\documentclass[12pt,letterpaper]{article}%
\usepackage{amsfonts}
\usepackage{amssymb}
\usepackage{amsmath}
\usepackage{amsthm}
\usepackage[round,authoryear]{natbib}
\usepackage{booktabs}
\usepackage{dcolumn}
\usepackage{graphicx,epstopdf}
\usepackage[T1]{fontenc}
\usepackage{lmodern}
\usepackage{enumitem}
\usepackage{subfig}
\usepackage{bm}
\usepackage[colorlinks,linkcolor=red,citecolor=blue,pdftex,breaklinks]%
{hyperref}
\usepackage[nameinlink]{cleveref}
\usepackage[flushleft]{threeparttable}
\usepackage{graphicx}%
\providecommand{\U}[1]{\protect\rule{.1in}{.1in}}
\hypersetup{bookmarksopen=true}
\Crefname{assumption}{Assumption}{Assumptions}

\makeatletter
\renewcommand*{\eqref}[1]{\hyperref[{#1}]{\textup{\tagform@{\ref*{#1}}}}}
\makeatother
\renewcommand{\baselinestretch}{1}
 \theoremstyle{plain}
\newtheorem{theorem}{Theorem}

\newtheorem{lemma}{Lemma}[section]
\theoremstyle{definition}
\newtheorem{assumption}{Assumption}
\newtheorem*{algorithm}{Algorithm for Implementing the LM Test Procedure}
\newtheorem*{algorithm1}{Algorithm for Implementing the Permutation Procedure}
\expandafter
\def \expandafter \normalsize \expandafter{\normalsize \setlength \abovedisplayskip{10pt plus 2pt minus 7pt}}
\expandafter
\def \expandafter \normalsize \expandafter{\normalsize \setlength \abovedisplayshortskip{0pt plus 2pt}}
\expandafter
\def \expandafter \normalsize \expandafter{\normalsize \setlength \belowdisplayskip{10pt plus 2pt minus 7pt}}
\expandafter
\def \expandafter \normalsize \expandafter{\normalsize \setlength \belowdisplayshortskip{5pt plus 2pt minus 3pt}}

\newcommand{\blind}{0}

\begin{document}

\def\spacingset#1{\renewcommand{\baselinestretch}%
{#1}\small\normalsize} \spacingset{1}

\if0\blind
{
  \title{\bf{Tests for Group-Specific Heterogeneity in High-Dimensional Factor
Models}\thanks{We are grateful for comments from Minsu Chang,
Silvia Goncalves, Doosoo Kim, Leo Michelis, Benoit Perron, and
seminar participants at the 2019 CIREQ Montreal Econometrics
Conference, the 2019 Annual Conference of the Canadian Economics
Association,  the 36th Meeting of the Canadian Econometric Study
Group, and Ryerson University. The authors gratefully acknowledge
financial support from the Social Sciences and Humanities Research
Council (SSHRC).}}
  \author{Antoine A. Djogbenou \hspace{.2cm}\\
    Department of Economics, York University\\
    and \\
    Razvan Sufana \\
    Department of Economics, York University}
  \maketitle
} \fi

\if1\blind
{
  \bigskip
  \bigskip
  \bigskip
  \begin{center}
    {\LARGE\bf Tests for Group-Specific Heterogeneity in High-Dimensional Factor Models}
\end{center}
  \medskip
} \fi


\begin{abstract}
Standard high-dimensional factor models assume that the comovements in a large set of variables
could be modeled using a small number of latent factors that affect all variables.
In many relevant applications in economics and
finance, heterogenous comovements specific to some known groups of variables
naturally arise, and reflect distinct cyclical movements within
those groups. This paper develops two new statistical tests
that can be used to investigate whether there is evidence
supporting group-specific heterogeneity in the data. The first test
statistic is designed for the alternative hypothesis of group-specific
heterogeneity appearing in at least one pair of groups; the second is
for the alternative of group-specific heterogeneity appearing in
all pairs of groups. We show that the second moment of factor loadings
changes across groups when heterogeneity is present, and use this feature to establish the
theoretical validity of the tests.
We also propose and prove the validity of a permutation approach for
approximating the asymptotic distributions of the two test statistics.
The simulations and the
empirical financial application indicate that the proposed tests
are useful for detecting group-specific heterogeneity.

\end{abstract}


\noindent%
\vfill

\newpage
\spacingset{1.45} 
\section{Introduction}

Large datasets of financial and macroeconomic variables are
currently easily accessible. In many policy-relevant applications,
empirical researchers have found it useful to summarize the
information in a high-dimensional set of economic variables by
extracting common underlying latent factors. As a consequence, the
modeling and extraction of underlying latent factors is important
for econometricians and practitioners alike. Factor models have
been widely used since the seminal paper of \citet*{SW2002} to
model the dynamics of the common factors, and the principal
components method (PCM) has remained a popular factor extraction
procedure. The PCM generally assumes a one-level structure, where
comovements in a large panel of variables are summarized by a few
latent factors affecting all variables. In particular, each
variable in a large panel can be decomposed into an idiosyncratic
error component and a common component (for details, see
\citet*{SW2002}; \citet*{BN2002}; \citet*{Bai2003}). The
popularity of the PCM as a factor extraction procedure relies on
its computational simplicity and its properties, established in
\citet*{Bai2003} and \citet*{BN2006}.

In many applications, such as asset pricing and studies of real
international economic activity, heterogeneous group structures
may naturally arise in the specification of the factor models. In
such cases, cyclical movements specific to some groups of
variables can be considered. 
In a study of real international economic
activity, \citet*{KOP2012} used these models to investigate the
decoupling of developed and emerging economy activity factors.
Although globalization has shaped the world economy in recent
decades, emerging economies have experienced impressive growth
relative to developed economies, suggesting the existence of
separate business cycles. Using a large set of observable economic
activity variables, \citet*{Djogbenou2019} proposed a statistical
test to investigate whether the latter assertion is supported by
observed data, and found evidence that suggests the existence of
specific comovement within the group of emerging economy variables
and the group of developed economy variables. However, this test
is only applicable to the case of two groups of variables.

This paper makes two contributions. The first is to extend
the analysis in \citet*{Djogbenou2019} to
situations where there may be heterogeneity specific to multiple
groups of variables.
We assume that the number of groups of variables and the group membership of each variable are known,
which is the case in certain real-world situations where these features are already predetermined,
such as the one considered in the application section.
These assumptions distinguish the framework of this paper from that used in \citet*{AB2017}, where
the number of groups is determined using a model selection criterion,
and the group membership of each variable is estimated from historical data.

We define the situations of group-specific heterogeneity that can occur,
some of which can only appear in settings with three or more groups.
We develop tests of the null hypothesis of no
group-specific heterogeneity against the alternative that
some form of
group-specific heterogeneity is present. Two types of
alternative hypotheses are considered. Under the first
alternative, group-specific heterogeneity emerges in at least one
pair of groups; under the second, group-specific heterogeneity emerges in
all pairs of groups. We propose two Lagrange Multiplier (LM) type
tests, derive their asymptotic distributions under the null
hypothesis, and show that each test statistic diverges under the
corresponding type of alternative hypothesis.
Compared to \citet*{Djogbenou2019}, where two groups are considered, our generalization to the multiple groups setting leads to some non-standard asymptotic distributions.
These asymptotic distributions are based on Chi-square random variables,
some of which are dependent, and can be simulated taking their dependence structure into account in order to
generate critical values for the tests.

The second contribution of the paper is to
consider and show the validity of a permutation approach for
approximating the asymptotic distributions of the two test statistics.
This approach has the advantage of avoiding the simulation of the asymptotic distributions.
Permutation tests are nonparametric tests that provide a superior control of the test size,
and are based on a standard exchangeability assumption on the distribution of the factor loadings under the null hypothesis.
For details regarding permutation inference, see, for example, \citet*{LR2005},
\citet*{CRS2017}, and \citet*{HG2018}.

Simulation experiments are used to study the finite sample performance of all tests.
We also consider an application in which we test for the existence of heterogeneous industry groups
in two classifications of industries in the U.S. financial sector. We use the Standard Industrial
Classification (SIC) to determine the groupings. The first
consists of six major groups within the Finance,
Insurance, and Real Estate division of the SIC, and the second
includes nine industry groups within the same division. The tests
reveal that there is group heterogeneity 
in all pairs of major groups.
For the second grouping, they show that
there is group heterogeneity in at least one pair of industry groups. 


The rest of the paper is organized as follows. In \Cref{setup}, we
present the model. In \Cref{asym_results}, we describe the test
statistics and the asymptotic validity results. \Cref{perm_results} presents the permutation versions of the tests. We investigate the finite sample properties of the proposed tests in \Cref{simulation}. An empirical illustration is included in
\Cref{application}. \Cref{conclusion} concludes the paper. Proofs are relegated to the Appendix. Throughout
the paper, the scalar $C$ denotes a generic constant, $\mathbb{I}\left(  \cdot\right)  $ is the indicator function and
$\left\Vert {\cdot}\right\Vert $ denotes the Euclidean norm. We also use the notation $\delta_{NT}=\min\left[
\sqrt{N},\sqrt{T}\right]  $.\bigskip

\section{The Model \label{setup}}

The goal of this paper is to develop tests for group-specific
heterogeneity in high-dimensional factor models. We assume the
underlying structure without  group-specific heterogeneity of a variable $i$ can be modeled using a large panel factor model given by%
\begin{equation}
X_{it}=\bm{\lambda}_{c,i}^{\prime}\bm{f}_{c,t}
+{e}_{it}, i=1,\ldots
,N,t=1,\ldots,T, \label{panel_model}%
\end{equation}
where $N$ is the number of variables, $T$ is the number of time
periods, and $t$ denotes the time period index. The vector
$\bm{f}_{c,t}$ of dimensions $r_{c}\times1$ contains common latent
factors affecting all variables. The factor loadings
$\bm{\lambda}_{c,i}:r_{c}\times1\ $ capture the exposure of the
variable $i$ to the common factors, and $\bm{e}_{it}$ denotes an
idiosyncratic error.

We assume there are $S$ groups and let the index $g_{i}\in\left\{
j,j=1,\ldots,S\right\} $ denote the group that includes the
variable $i$. We let $N_{j}$ denote the number of variables in
group $j$ such that $N=\sum_{j=1}^{S}N_{j}$, and let
$M_{j}=\sum_{l=1}^{j}N_{l},$ $j=1,\ldots,S,$ with $M_{0}=0$. We
also assume that for any $j,$ $\frac{N_{j}}{N}\ $converges to a
fixed number $\pi_{j}\in\left[  \alpha_{1},\alpha_{2}\right]
\subset\left(  0,1\right) $.

The factor model can be written in matrix form as%
\begin{equation}
{\bm{X}}=\bm\Lambda\bm F_{c}^{\prime}+\bm{e}=\left[
\bm\Lambda_{1}^{\prime }\cdots\bm\Lambda_{S}^{\prime}\right]
^{\prime}\bm F_{c}^{\prime}+\bm{e},
\end{equation}
where ${\bm{X}=}\left(  X_{it}\right)  :N\times T$ is the matrix
of observed variables, $\bm F_{c}^{\prime}=\left[
\bm{f}_{c,1}\cdots\bm{f}_{c,T}\right] :r_{c}\times T$ contains the
common factors, ${\bm{e}=}\left(  e_{it}\right) :N\times T$
collects the idiosyncratic errors,
\[
\bm\Lambda=\left[
\begin{array}
[c]{c}%
\bm\lambda_{c,1}^{\prime}\\
\vdots\\
\bm\lambda_{c,N}^{\prime}%
\end{array}
\right]  :N\times r_{c}
\]
is the matrix of factor loadings, and
\[
\bm\Lambda_{j}=\left[
\begin{array}
[c]{c}%
\bm\lambda_{c,(M_{j-1}+1)}^{\prime}\\
\vdots\\
\bm\lambda_{c,M_{j}}^{\prime}%
\end{array}
\right]  :N_{j}\times r_{c},j=1,\ldots,S.
\]
captures the exposure of variables in group $j$ to the factors in
$\bm F_{c}$.
%

We impose the following assumptions, where we let $\iota_{ji}\equiv\mathbb{I}\left(
M_{j-1}+1\leq i\leq M_{j}\right)  ,$ with $j=1,\ldots,S,$ and $i=1,\ldots
,N.$

\begin{assumption}
\label{Ass1}(Factor model and idiosyncratic errors)
\end{assumption}

\begin{description}
\item[(a)] $\mathrm{E}\left\Vert \bm{f}_{c,t}\right\Vert ^{4}\leq C$ and
$\frac{1}{T}\bm{F}_{c}^{\prime}\bm{F}_{c}=\frac{1}{T}\sum_{t=1}^{T}%
\bm{f}_{c,t}\bm{f}_{c,t}^{\prime}\overset{P}{\longrightarrow}%
\bm{\Sigma }_{\bm{F}}>0$, where $\bm{\Sigma }_{\bm{F}}$ is non-random.

\item[(b)] The factor loadings $\left\{  \bm{\lambda}_{c,i}\right\}
_{i=1,\ldots,N}$ are independent across $i,$ and $\mathrm{E}\left(
\bm{\lambda}_{c,i}\bm{\lambda}_{c,i}^{\prime}\right)  =\bm{\Sigma
}_{\bm{\Lambda}}>\bm0$, where $\bm{\Sigma }_{\bm{\Lambda}}$ is non-random,
such that \textrm{E}$\left(  \left\Vert \bm{\lambda}_{c,i}\right\Vert ^{8
}\right)  \leq C$.

\item[(c)] The eigenvalues of the $r\times r$ matrix $\left(
\bm{\Sigma }_{\bm{F}}\times\bm{\Sigma }_{\bm{\Lambda}}\right)  $ are distinct.

\item[(d)] $\mathrm{E}\left(  e_{it}\right)  =0,$ $\mathrm{E}\left\vert
e_{it}\right\vert ^{8}\leq C.$

\item[(e)] $\mathrm{E}\left(  e_{it}e_{js}\right)  =\sigma_{ij,ts}%
$,\ $\left\vert \sigma_{ij,ts}\right\vert \leq\overline{\sigma}_{ij}$ for all
$\left(  t,s\right)  $ and $\left\vert \sigma_{ij,ts}\right\vert \leq\tau
_{st}$ for all $\left(  i,j\right)  $, with $\frac{1}{N}\sum_{i,j=1}%
^{N}\overline{\sigma}_{ij}\leq C,$ $\frac{1}{T}\sum_{t,s=1}^{T}\tau_{st}\leq
C$ and $\frac{1}{NT}\sum_{i,j,t,s=1}\left\vert \sigma_{ij,ts}\right\vert \leq
C.$

\item[(f)] $\mathrm{E}\left\vert \frac{1}{\sqrt{N}}\sum_{i=1}^{N}\left(
e_{it}e_{is}-E\left(  e_{it}e_{is}\right)  \right)  \right\vert ^{4}\leq C$
for all $\left(  t,s\right)  .$

\item[(g)] The limit $\bm{S}_{0}$ of $\mathrm{Var}\left(  \sqrt{\frac{N_{j}%
}{N}}\bm{A}_{0}\left(  j,\bm{\Lambda}\bm{H}_{0}^{\prime-1}\right)  \right)  $
is bounded and positive definite, where%
\[
\bm{A}_{0}\left(  j,\bm{\Lambda}\bm{H}_{0}^{\prime-1}\right)  =\mathrm{Vech}%
\left(  \frac{\sqrt{N}}{N_{j}}\sum_{i=M_{j-1}+1}^{M_{j}}\bm{H}_{0}^{-1}\left(
\bm{\lambda}_{c,i}\bm{\lambda}_{c,i}^{\prime}-\bm{\Sigma
}_{\bm{\Lambda}}\right)  \bm{H}_{0}^{-1}\right)  ,
\]
and $\bm{H}_{0}$ is defined in \Cref{estimation}.

\item[(h)] \textrm{E}$\left(  \left(  \mathrm{Vech}\left(  \bm{H}_{0}%
^{-1}\bm{\lambda}_{c,i}\bm{\lambda}_{c,i}^{\prime}\bm{H}_{0}^{\prime
-1}\right)  \right)  \mathrm{Vech}\left(  \bm{H}_{0}^{-1}\bm{\lambda}_{c,i}%
\bm{\lambda}_{c,i}^{\prime}\bm{H}_{0}^{\prime-1}\right)  ^{\prime}\right)
=\bm{\Sigma}_{\bm{\Lambda}\bm{\Lambda}}$.
\end{description}

\begin{assumption}
\label{Ass2}(Moment conditions and weak dependence among $\{\bm{f}_{c,t}\},$
$\{\bm{\lambda}_{c,i}\}$ and $\{e_{it}\}$)
\end{assumption}

\begin{description}
\item[(a)] $\mathrm{E}\left(  \frac{1}{N}\sum_{i=1}^{N}\left\Vert \frac
{1}{\sqrt{T}}\sum_{t=1}^{T}\bm{f}_{c,t}e_{it}\iota_{mi}\right\Vert
^{2}\right)  \leq C$, $m=1,\ldots,S$, where $\mathrm{E}\left(  \bm{f}_{c,t}%
e_{it}\right)  =\bm 0$ for every $\left(  i,t\right)  $.

\item[(b)] For each $t,$ $\mathrm{E}\left\Vert \frac{1}{\sqrt{TN}}\sum
_{s=1}^{T}\sum_{i=1}^{N}\bm{f}_{c,s}\left(  e_{it}e_{is}-\mathrm{E}\left(
e_{it}e_{is}\right)  \right)  \right\Vert ^{2}\leq C$.

\item[(c)] $\mathrm{E}\left\Vert \frac{1}{\sqrt{TN}}\sum_{t=1}^{T}\sum
_{i=1}^{N}\bm{f}_{c,t}\bm{\lambda}_{c,i}^{\prime}e_{it}\iota_{mi}\right\Vert
^{2}\leq C$, $m=1,\ldots,S$, where $\mathrm{E}\left(  \bm{f}_{c,t}%
\bm{\lambda}_{c,i}^{\prime}e_{it}\right)  =\bm0$ for all $\left(  i,t\right)
$.

\item[(d)] $\mathrm{E}\left(  \frac{1}{T}\sum_{t=1}^{T}\left\Vert \frac
{1}{\sqrt{N}}\sum_{i=1}^{N}\bm{\lambda}_{c,i}e_{it}\iota_{mi}\right\Vert
^{2}\right)  \leq C$, $m=1,\ldots,S$, where $\mathrm{E}\left(
\bm{\lambda}_{c,i}e_{it}\right)  =\bm{0}$ for all $\left(  i,t\right)  $.
\end{description}

\Cref{Ass1,Ass2} allow for weak dependence and heteroskedasticity
in the idiosyncratic errors and are similar to the assumptions
A$-$D of \citet*{BN2002}, 1$-$3 of \citet*{DGP2015} and 1$-$2 of
\citet*{Djogbenou2019b}. \Cref{Ass1} (b), (g) and (h) are
useful for deriving the asymptotic distribution of our proposed
test statistics under the null hypothesis and impose the existence
of slightly more than the fourth moment of $\bm{\lambda}_{c,i}$.
Using a law of large numbers, \Cref{Ass1} (b) implies a standard
assumption for factor models, stating that
$\frac{1}{N}\bm{\Lambda}^{\prime}\bm{\Lambda}=\frac{1}{N}\sum_{i=1}%
^{N}\bm{\lambda}_{c,i}\bm{\lambda}_{c,i}^{\prime}\overset{P}{\longrightarrow
}\bm{\Sigma }_{\bm{\Lambda}}>0$, where $\bm{\Sigma
}_{\bm{\Lambda}}$ is non-random. Moreover, \Cref{Ass2} (a), (c)
and (d), which restricts the dependence between $\bm{f}_{c,t}$,
$\bm{\lambda}_{c,i}$ and $e_{it}$ among specific groups of
variables, is slightly stronger.

In the model presented thus far, the only factors driving the
variables are common factors. In this case, the exposure of the
variables to the factors is homogenous across groups, and all
variables share common cyclical movements. However, in the
presence of group-specific heterogeneity, we expect some
factors to appear and affect the groups differently, implying
distinct cyclical movements across the groups of variables.

As shown in the next section, the second moment of factor loadings
changes in presence of group-specific heterogeneity. Nevertheless,
under the null hypothesis $H_{0}$ of no group-specific
heterogeneity, the scaled product of
the factor loadings for groups $j$ is%
\[
\frac{1}{N_{j}}\bm{\Lambda}_{j}^{\prime}\bm{\Lambda}_{j}=\frac{1}{N_{j}}%
\sum_{i=M_{j-1}+1}^{M_{j}}\bm\lambda_{c,i}\bm\lambda_{c,i}^{\prime}%
=\bm\Sigma_{\bm{\Lambda}}+o_{P}(1),
\]
under \Cref{Ass1}(b), and therefore the
difference between
the scaled products for groups $j$ and $k$ ($j<k$) is%
\[
\frac{1}{N_{j}}\bm{\Lambda}_{j}^{\prime}\bm{\Lambda}_{j}-\frac{1}{N_{k}%
}\bm{\Lambda}_{k}^{\prime}\bm{\Lambda}_{k}=o_{P}(1).
\]
We distinguish two types of alternative hypotheses. The first, denoted $H_{1}%
$, corresponds to the case where group-specific heterogeneity
emerges in at least one pair of groups. The second, denoted $H_{2}$,
corresponds to the case where group-specific heterogeneity emerges
in all pairs of groups. Given the model, the hypothesis $H_{2}$
also refers to the case where there is no pair of groups without
group-specific heterogeneity.
For convenient reference, we list these hypotheses together:

$H_{1}$: Group-specific heterogeneity emerges in at least one pair of groups.

$H_{2}$: Group-specific heterogeneity emerges in all pairs of groups.\\
\Cref{alternative} formally presents
the model when
different types of
group heterogeneity are present.

\subsection{Group-specific heterogeneity\label{alternative}}

Group heterogeneity in factor models may arise in the form of
group-specific factors.
In this case, the specification of the variable $i$ is
\begin{equation}
X_{it}=\bm{\lambda}_{c,i}^{\prime}\bm{f}_{c,t}+\bm{\lambda}_{g_{i},i}^{\prime
}\bm{f}_{g_{i},t}+{e}_{it},i=1,\ldots,N,t=1,\ldots,T. \label{panel_model_h1}%
\end{equation}
The vector $\bm{f}_{g_{i},t}$ of dimensions $r_{g_{i}}\times1$
includes
latent specific factors affecting
only the variables in group $g_{i}.$ The factor loadings
$\bm{\lambda}_{g_{i},i}:r_{g_{i}}\times1$ capture the exposure of
the variable $i$ to the specific factors in group $g_{i}$.

The matrix equivalent of \Cref{panel_model_h1} is
\begin{equation}
{\bm{X}}=\bm\Phi\bm G^{\prime}+\bm{e}=\left[  \bm\Phi_{1}^{\prime}%
\cdots\bm\Phi_{S}^{\prime}\right]  ^{\prime}\left[  \bm F_{c}\text{  }\bm F_{1}%
\cdots\bm F_{S}\right]  ^{\prime}+\bm{e}, \label{one_level_repr2}%
\end{equation}
where $\bm F_{j}^{\prime}=\left[
\bm{f}_{j,1}\cdots\bm{f}_{j,T}\right] :r_{j}\times T$ is the
matrix of factors specific to group $j$, $j=1,\ldots ,S$, and the
matrices
\[
\bm\Phi_{j}=\left[
\bm\phi_{M_{j-1}+1}\cdots\bm\phi_{M_{j}}\right]  ^{\prime }=\left[
\begin{array}
[c]{cccc}%
\bm\lambda_{c,M_{j-1}+1}^{\prime} & \bm0 & \bm\lambda_{j,
M_{j-1}+1 }^{\prime}
& \bm0\\
\vdots & \vdots & \vdots & \vdots\\
\bm\lambda_{c,M_{j}}^{\prime} & \bm0 &
\bm\lambda_{j,M_{j}}^{\prime} & \bm0
\end{array}
\right]  :N_{j}\times r,j=1,\ldots,S,
\]
with $r=r_{c}+r_{1}+\ldots+r_{S},$ reflect the exposure to the
latent factors in the factor model representation in
\Cref{one_level_repr2}. The first $r_{c}$ columns contain the
factor loadings that capture the exposure of the
variables in group $j$ to the common factors. The following $\sum_{l=1}%
^{j-1}r_{l}$ columns and the last $\sum_{l=j+1}^{S}r_{l}$ columns
have only zero elements because the variables in group $j$ have
zero exposure to the specific factors $\bm F_{1},\ldots,\bm
F_{j-1},\bm F_{j+1},\ldots,\bm F_{S}.$ The columns of factors
loadings between the zero columns reflect how the variables in
group $j$ react to specific factors in $\bm F_{j}$.

\begin{assumption}
\label{Ass3} (Additional conditions for group heterogeneous factor models)
\end{assumption}

\begin{description}
\item[(a)] $\frac{1}{T}\bm{G}^{\prime}\bm{G}=\frac{1}{T}\sum_{t=1}%
^{T}\bm{g}_{t}\bm{g}_{t}^{\prime}\overset{P}{\longrightarrow}%
\bm{\Sigma }_{\bm{G}}>\bm{0}$, where $\bm{\Sigma }_{\bm{G}}$ is non-random and
$\bm{g}_{t}^{\prime}$
denotes
the $t^{th}$ row of the matrix $\bm G$.

\item[(b)] The factor loadings $\left\{  \bm{\phi}_{i}\right\}  _{i=1,\ldots
,N}$ are independent across $i$, \textrm{E}$\left(  \left\Vert \bm{\phi}_{i}%
\right\Vert ^{4}\right)  \leq C$, and for $i$ in group $j,$ $\mathrm{E}\left(
\bm{\phi}_{i}\bm{\phi}_{i}^{\prime}\right)  $ is the limit defined in
\Cref{limit} or \Cref{limit_h2},
such that$\frac{1}{N}\bm{\Phi}^{\prime}\bm{\Phi}=\frac{1}{N}\sum_{i=1}%
^{N}\bm{\phi}_{i}\bm{\phi}_{i}^{\prime}\overset{P}{\longrightarrow
}\bm{\Sigma }_{\bm{\Phi}}>\bm{0}$, where $\bm{\Sigma }_{\bm{\Phi}}$ is non-random.

\item[(c)] The eigenvalues of the matrix $\bm{\Sigma
}_{\bm{G}}\times\bm{\Sigma }_{\bm{\Phi}}$ are distinct.

\item[(d)] $\mathrm{E}\left(  \frac{1}{N}\sum_{i=1}^{N}\left\Vert \frac
{1}{\sqrt{T}}\sum_{t=1}^{T}\bm{g}_{t}e_{it}\iota_{ji}\right\Vert ^{2}\right)
\leq C$, $j=1,\ldots,S$, where $\mathrm{E}\left(  \bm{g}_{t}e_{it}\right)
=\bm0$ and $\mathrm{E}\left(  \bm{g}_{t}\bm{\phi}_{g_{i},i}^{\prime}%
e_{it}\right)  =\bm{0}$\ for every $\left(  i,t\right)  $.

\item[(e)] For each $t,$ $\mathrm{E}\left\Vert \frac{1}{\sqrt{TN}}\sum
_{s=1}^{T}\sum_{i=1}^{N}\bm{g}_{s}\left(  e_{it}e_{is}-\mathrm{E}\left(
e_{it}e_{is}\right)  \right)  \right\Vert ^{2}\leq C$.

\item[(f)] $\mathrm{E}\left\Vert \frac{1}{\sqrt{TN}}\sum_{t=1}^{T}\sum
_{i=1}^{N}\bm{g}_{t}\bm{\phi}_{g_{i},i}^{\prime}e_{it}\iota_{ji}\right\Vert
^{2}\leq C$, $j=1,\ldots,S$.

\item[(g)] $\mathrm{E}\left\Vert \frac{1}{\sqrt{TN}}\sum_{t=1}^{T}\sum
_{i=1}^{N}\bm{g}_{t}\bm{\phi}_{g_{i},i}^{\prime}e_{it}\iota_{ji}\right\Vert
^{2}\leq C$, $j=1,\ldots,S$, where $\mathrm{E}\left(  \bm{g}_{t}%
\bm{\phi}_{g_{i},i}^{\prime}e_{it}\right)  =\bm0$ for all $\left(  i,t\right)
$.

\item[(h)] For any $j\neq k,$ the limit in probability $\bm{S}_0(j,k)$ of $\bm{S}\left(
j,k,\bm\Phi\bm{\Xi}_{0}^{\prime-1}\right)  $ is positive definite,
where $\bm{\Xi}_{0}$ is defined in \Cref{estimation} and $\bm{S}$ is defined in \Cref{asym_results}.


\end{description}


 \Cref{Ass3} (a)$-$(g) complements the previous assumptions for the factor
model representation in \Cref{one_level_repr2} in the presence of group-specific heterogeneity.
 \Cref{Ass3} (h) imposes positive definiteness on the limit of the variance estimator under the alternative.

Let us now consider a pair of groups $(j,k)$ ($j<k$), such that
group-specific factors emerge in group $j$ or group $k$.
According to \Cref{Ass3} (b), the scaled product of the factor
loadings for
group $j$, $\frac{1}{N_{j}}\bm{\Phi}_{j}^{\prime}\bm{\Phi}_{j}$, is%
\begin{equation}
\left[
\begin{array}
[c]{cccc}%
\frac{1}{N_{j}}\sum_{i=M_{j-1}+1}^{M_{j}}\bm\lambda_{c,i}\bm\lambda
_{c,i}^{\prime} & \bm0 & \frac{1}{N_{j}}\sum_{i=M_{j-1}+1}^{M_{j}}%
\bm\lambda_{c,i}\bm\lambda_{j,i}^{\prime} & \bm0\\
\bm0 & \bm0 & \bm0 & \bm0\\
\frac{1}{N_{j}}\sum_{i=M_{j-1}+1}^{M_{j}}\bm\lambda_{j,i}\bm\lambda
_{c,i}^{\prime} & \bm0 & \frac{1}{N_{j}}\sum_{i=M_{j-1}+1}^{M_{j}}%
\bm\lambda_{j,i}\bm\lambda_{j,i}^{\prime} & \bm0\\
\bm0 & \bm0 & \bm0 & \bm0
\end{array}
\right]  =\left[
\begin{array}
[c]{cccc}%
\bm\Sigma_{\bm{\Lambda}} & \bm0 & \bm\Sigma_{cj} & \bm0\\
\bm0 & \bm0 & \bm0 & \bm0\\
\bm\Sigma_{cj}^{\prime} & \bm0 & \bm\Sigma_{jj} & \bm0\\
\bm0 & \bm0 & \bm0 & \bm0
\end{array}
\right]  +o_{P}(1), \label{limit}%
\end{equation}
where $\bm\Sigma_{\bm{\Lambda}}\equiv\mathrm{E}\left(  \bm\lambda
_{c,i}\bm\lambda_{c,i}^{\prime}\right)
,\bm\Sigma_{jj}\equiv\mathrm{E}\left(
\bm\lambda_{j,i}\bm\lambda_{j,i}^{\prime}\right)  ,$ and $\bm\Sigma_{cj}%
\equiv\mathrm{E}\left(
\bm\lambda_{c,i}\bm\lambda_{j,i}^{\prime}\right) $. Then
$\bm\Sigma_{jj}\neq\bm0$ or $\bm\Sigma_{kk}\neq\bm0$ since the
factor loadings associated with specific factors within groups $j$
and $k$ would be $\bm0$ otherwise. Therefore, the difference
between the scaled products for groups $j$ and $k$
\begin{equation}
\frac{1}{N_{j}}\bm{\Phi}_{j}^{\prime}\bm{\Phi}_{j}-\frac{1}{N_{k}%
}\bm{\Phi}_{k}^{\prime}\bm{\Phi}_{k}, \label{limit_alternative}%
\end{equation}
converges in probability to a nonzero matrix. When the difference
in \eqref{limit_alternative} is scaled by $\sqrt{N},$\ it
diverges. As described in \Cref{asym_results}, we construct a test statistic
which diverges when at least one of the groups $j$ and $k$ is
heterogeneous, that is, contains group-specific factors.

\bigskip

Group heterogeneity in factor models may also arise in the form of
factors affecting the variables within groups differently
and which can be identified through distinct second moments of loadings
or distinct cross moments between loadings and global loadings.
In order to formally express this situation,
let us assume that a factor $\bm F_{s}:T\times r_{s}$ affects
both groups in some pair of groups $(j,k)$ ($j<k$).
For simplicity of exposition, suppose the first type of heterogeneity discussed above is not also present.
The matrix notation in \Cref{one_level_repr2} becomes
$\bm G=\left[ \bm F_{c}\text{ }\bm F_{s}\right]:T\times(r_{c}+r_{s})$
and the factor loadings of the variables in group $j$ are
\begin{equation}
\bm\Phi_{j}=\left[ \bm\phi_{M_{j-1}+1}\cdots\bm\phi_{M_{j}}\right]
^{\prime }=\left[
\begin{array}
[c]{cc}%
\bm\lambda_{c,M_{j-1}+1}^{\prime} &
\bm\lambda_{j,M_{j-1}+1}^{\prime}\\
\vdots & \vdots\\
\bm\lambda_{c,M_{j}}^{\prime} & \bm\lambda_{j,M_{j}}^{\prime}
\end{array}
\right] : N_{j} \times(r_{c}+r_{s}),
\end{equation}
\noindent which leads to
\begin{equation}
\frac{1}{N_{j}}\bm{\Phi}_{j}^{\prime}\bm{\Phi}_{j}=\left[
\begin{array}
[c]{cc}%
\bm\Sigma_{\bm{\Lambda}} & \bm\Sigma_{cj}\\
\bm\Sigma_{cj}^{\prime} & \bm\Sigma_{jj}
\end{array}
\right]  +o_{P}(1).\label{limit_h2}%
\end{equation}
$\bm\Phi_{k}$ and $\frac{1}{N_{k}}\bm{\Phi}_{k}^{\prime}\bm{\Phi}_{k}$
can be expressed in a similar way.
The difference between the scaled products for the
groups in the pair $(j,k)$ is
\begin{equation}
\frac{1}{N_{j}}\bm{\Phi}_{j}^{\prime}\bm{\Phi}_{j}-\frac{1}{N_{k}%
}\bm{\Phi}_{k}^{\prime}\bm{\Phi}_{k}=\left[
\begin{array}
[c]{cc}%
\bm0 & \bm\Sigma_{cj}-\bm\Sigma_{ck}\\
\bm\Sigma_{cj}^{\prime}-\bm\Sigma_{ck}^{\prime}
& \bm\Sigma_{jj}-\bm\Sigma_{kk}
\end{array}
\right]  +o_{P}(1).\label{limit_alternative_h2}%
\end{equation}
Heterogeneity in the pair of groups $(j,k)$ arises if $\bm\Sigma_{jj} \ne \bm\Sigma_{kk}$
or $\Sigma_{cj} \ne \Sigma_{ck}$. When this occurs, the difference in \eqref{limit_alternative_h2} converges in probability to a nonzero matrix, which diverges when scaled by $\sqrt{N}$.

To summarize, the first type of group-specific heterogeneity is induced by some factors
affecting only the variables in each individual group,
while the second type of group-specific heterogeneity
is induced by some factors affecting the groups of variables in different ways.
Note that a combination of both types of heterogeneity can be present at the same time.
The proposed statistical tests developed in the next section can be used to detect group-specific heterogeneity of either type discussed above.

\subsection{Model estimation\label{estimation}}

In practice, the factors and factor loadings are latent and need
to be estimated. A popular approach consists of using the PCM to
obtain
$\hat{\bm\Lambda}=\left[  \hat{\bm\lambda}_{1}\cdots\hat{\bm\lambda}%
_{N}\right]  ^{\prime}$\ and $\hat{\bm F}=\left[  \hat{\bm
f}_{1}\cdots\hat{\bm f}_{T}\right]  ^{\prime},$ which represent
the estimates of $\bm\Lambda$ and $\bm F_{c}$ when there is no
group heterogeneity ($\bm\Phi$ and $\bm G$ when group
heterogeneity is present). These estimates can be obtained by
minimizing the sum of squared idiosyncratic residuals under the
restriction $\hat{\bm\Lambda}^{\prime}\hat{\bm\Lambda}/N=\bm I$.
The estimated factor loadings are given by $\sqrt{N}$ times the
$r$ eigenvectors associated with the $r$ largest eigenvalues of
$\bm X\bm X^{\prime}$, while the estimated factors are given by
$\bm X^{\prime}\hat{\bm\Lambda}/N$ (see Bai and Ng (2008)).
We use the information criteria suggested by Bai and Ng (2002)
to select the number of latent factors and let this number be
$r\geq r_{c}$ in general. As shown in Bai and Ng (2002)
and Bai (2003),
at any time period $t,$ $\hat{\bm f}_{t}$ consistently estimates a
rotation of the true factor space, given by
$\hat{\bm{H}}^{\prime}\bm f_{c,t}$ under no group heterogeneity
and $\hat{\bm{\Xi}}^{\prime}\bm g_{t}$ under group heterogeneity,
where $g_{t}^{\prime}$ is a typical row of $\bm G,$ and
$\hat{\bm{H}}:r_{c}\times r_{c}$ and $\hat{\bm{\Xi}}:r\times r$
are the
rotation matrices$.$ In the former case, we have%
\begin{equation}
\bm{X}=\bm{\Lambda}\bm{F}_{c}^{\prime}+\bm{e}=\left(
\bm{\Lambda}\hat {\bm{H}}^{\prime-1}\right)  \left(
\bm{F}_{c}\hat{\bm{H}}\right)  ^{\prime
}+\bm{e}, \label{glo}%
\end{equation}
and in the latter case%
\begin{equation}
\bm{X}=\bm{\Phi}\bm{G}^{\prime}+\bm{e}=\left(  \bm{\Phi}\hat{\bm{\Xi}}%
^{\prime-1}\right)  \left(  \bm{G}\hat{\bm{\Xi}}\right)
^{\prime}+\bm{e}.
\label{spe}%
\end{equation}
The factor loading estimates $\hat{\bm{\Lambda}}$ and
$\hat{\bm{\Phi}}$ in \Cref{glo} and \Cref{spe} converge to their
rotated versions $\bm{\Lambda}\hat {\bm{H}}^{\prime-1}$ and
$\bm{\Phi}\hat{\bm{\Xi}}^{\prime-1}$, respectively. The
invertible limits of $\hat{\bm{H}}$ and $\hat{\bm{\Xi}}$ will be
denoted $\bm{H}_{0}$ and $\bm{\Xi}_{0},$ respectively.

We consider two
LM-type test statistics for each alternative hypothesis. The proposed testing procedures have the
advantage that the estimation of the group-specific factors, which
may be computationally intensive, is not required.\bigskip

\section{Test Statistics and Asymptotic Validity Results \label{asym_results}}

In this section, we describe the test statistics and establish
their formal validity. We consider two types of LM statistics that
correspond to the alternative hypotheses $H_{1}$\ and $H_{2},$
derive their limiting distributions under the null hypothesis, and
show that they tend to infinity under the corresponding
alternative hypothesis. To construct the statistics, we consider
the pair of groups $\left(  j,k\right)  $ and replace the latent
factor loadings in \Cref{limit_alternative} by their estimated
version to have
the scaled difference%
\begin{equation}
\bm{A}\left(  j,k,\hat{\bm\Lambda}\right)  =\mathrm{Vech}\left(
\sqrt
{N}\left(  \frac{1}{N_{j}}\sum_{i=M_{j-1}+1}^{M_{j}}\hat{\bm\lambda}_{i}%
\hat{\bm\lambda}_{i}^{\prime}-\frac{1}{N_{k}}\sum_{i=M_{k-1}+1}^{M_{k}}%
\hat{\bm\lambda}_{i}\hat{\bm\lambda}_{i}^{\prime}\right)  \right)
.
\label{test_A}%
\end{equation}
If group-specific heterogeneity emerges in the pair of groups $(j,k)$,
Appendix A shows that the difference
\begin{equation}
\frac{1}{N_{j}}\sum_{i=M_{j-1}+1}^{M_{j}}\hat{\bm\lambda}_{i}\hat{\bm\lambda
}_{i}^{\prime}-\frac{1}{N_{k}}\sum_{i=M_{k-1}+1}^{M_{k}}\hat{\bm\lambda}%
_{i}\hat{\bm\lambda}_{i}^{\prime}\overset{P}{\longrightarrow}
{\bm{\Xi}}^{-1}_{0}\mathrm{plim}\left(
\frac{1}{N_{j}}\bm{\Phi}_{j}^{\prime
}\bm{\Phi}_{j}-\frac{1}{N_{k}}\bm{\Phi}_{k}^{\prime}\bm{\Phi}_{k}\right)
\bm{\Xi}_{0}^{^{\prime}-1}\equiv \bm R_0(j,k),
\end{equation}
where $\bm{\Xi}_{0}$ is positive definite and $\mathrm{plim}
\left(  \frac
{1}{N_{j}}\bm{\Phi}_{j}^{\prime}\bm{\Phi}_{j}-\frac{1}{N_{k}}\bm{\Phi}_{k}%
^{\prime}\bm{\Phi}_{k}\right)  $ is nonzero as explained in
Section 2. This
implies that $\frac{1}{N_{j}}\sum_{i=M_{j-1}+1}^{M_{j}}\hat{\bm\lambda}%
_{i}\hat{\bm\lambda}_{i}^{\prime}-\frac{1}{N_{k}}\sum_{i=M_{k-1}+1}^{M_{k}%
}\hat{\bm\lambda}_{i}\hat{\bm\lambda}_{i}^{\prime}$ has a nonzero
limit. Thus, \Cref{test_A} diverges as the sample sizes increase.

Under the null hypothesis,
group-specific heterogeneity does not emerge in any pair $(j,k)$, and using a
decomposition similar
to the one in \Cref{decomp_alt}, we have%
\begin{equation}
\frac{1}{N_{j}}\sum_{i=M_{j-1}+1}^{M_{j}}\hat{\bm\lambda}_{i}\hat{\bm\lambda
}_{i}^{\prime}-\frac{1}{N_{k}}\sum_{i=M_{k-1}+1}^{M_{k}}\hat{\bm\lambda}%
_{i}\hat{\bm\lambda}_{i}^{\prime}\overset{P}{\longrightarrow}
{\bm{H}}^{-1}_{0}\mathrm{plim}\left(
\frac{1}{N_{j}}\bm{\Lambda}_{j}^{\prime
}\bm{\Lambda}_{j}-\frac{1}{N_{k}}\bm{\Lambda}_{k}^{\prime}\bm{\Lambda}_{k}%
\right)  \bm{H}_{0}^{^{\prime}-1}=\bm{0}.
\end{equation}
In this case, we show that the expression in \Cref{test_A} no
longer diverges.

For two groups $j$ and $k$,
we normalize $\bm{A}\left(
j,k,\hat{\bm\Lambda}\right)  $ using its variance estimator%
\[
\bm{S}\left(  j,k,\hat{\bm\Lambda}\right)  =\left(
\frac{N}{N_{j}}+\frac {N}{N_{k}}\right)
\frac{1}{N}\sum_{i=1}^{N}\mathrm{Vech}\left(
\hat{\bm\lambda}_{i}\hat{\bm\lambda}_{i}^{\prime}-\bm{I}\right)
\mathrm{Vech}\left(
\hat{\bm\lambda}_{i}\hat{\bm\lambda}_{i}^{\prime
}-\bm{I}\right)  ^{\prime}%
\]
and consider the LM-type statistic given by%
\begin{equation}
LM_{N}\left(  j,k,\hat{\bm\Lambda}\right)  =\bm{A}\left(
j,k,\hat{\bm\Lambda }\right)  ^{\prime}\left(  \bm{S}\left(
j,k,\hat{\bm\Lambda}\right)  \right)
^{-1}\bm{A}\left(  j,k,\hat{\bm\Lambda}\right)  . \label{test_stat}%
\end{equation}
We establish under \Cref{Ass1,Ass2} that\ $LM_{N}\left(
j,k,\hat{\bm\Lambda }\right)  $ follows asymptotically a
Chi-square distribution with $\frac{r\left(  r+1\right)  }{2}$
degrees of freedom.\ To ensure that the suggested test statistic
will diverge under a group heterogeneity alternative,
we propose two different test statistics given by%
\begin{equation}
LM_{1N}\left(  \hat{\bm\Lambda}\right)  =\max_{1\leq j<k\leq
S}LM_{N}\left( j,k,\hat{\bm\Lambda}\right)  ,
\end{equation}
when testing for $H_{1},$\ and
\begin{equation}
LM_{2N}\left(  \hat{\bm\Lambda}\right)  =\min_{1\leq j<k\leq
S}LM_{N}\left( j,k,\hat{\bm\Lambda}\right)  ,
\end{equation}
when testing for $H_{2}$. The following theorem derives the
limiting distributions of these two test statistics under the null
hypothesis.

%

\begin{theorem}
\label{Thm1}Suppose that \Cref{Ass1,Ass2} are satisfied. As
$N,T\rightarrow
\infty,$ if $\sqrt{N}/T\rightarrow0,$ then under the null, it holds that%
\begin{equation}
LM_{1N}\left(  \hat{\bm\Lambda}\right)  \overset{d}{\longrightarrow}%
\max_{1\leq j<k\leq S}
Q_{j,k}  , \label{Asym1}%
\end{equation}
and%
\begin{equation}
LM_{2N}\left(  \hat{\bm\Lambda}\right)  \overset{d}{\longrightarrow}%
\min_{1\leq j<k\leq S}
Q_{j,k}  , \label{Asym2}%
\end{equation}
where
\begin{equation}
\label{Z}
Q_{j,k}= \left(  \pi_{j}^{-1/2}\bm{Z}_{j}-\pi_{k}%
^{-1/2}\bm{Z}_{k}\right)  ^{\prime}\left(  \pi_{j}^{-1}+\pi_{k}^{-1}\right)
^{-1}\left(  \pi_{j}^{-1/2}\bm{Z}_{j}-\pi_{k}^{-1/2}\bm{Z}_{k}\right),
\end{equation}
for $1\leq j<k\leq S$,
with $\bm{Z}_{j}\sim\mathrm{N}\left(
\bm{0}_{r_{c}\left(  r_{c}+1\right)  /2},\bm{I}_{r_{c}\left(  r_{c}+1\right)
/2}\right)$,
$j=1,\ldots,S$,
are $\frac{S(S-1)}{2}$ random variables with a
Chi-square distribution with $\frac{r_{c}\left( r_{c}+1\right)
}{2}$ degrees of freedom.
\end{theorem}

The proof of \Cref{Thm1} is in Appendix A.
\Cref{Thm1} suggests that we could test the null hypothesis using
critical values from the derived asymptotic distributions under the stated assumptions.
These distributions are based on
$\frac{S\left(  S-1\right)  }{2}$ Chi-square random variables,
some of which are dependent as shown in \Cref{Z}.
The Chi-square variables can be simulated taking their dependence into account by generating the normal vectors $\bm{Z}_{j}$, $j=1,\ldots,S$,
and using the estimators $\frac{N_{j}}{N} $ of $\pi_{j}$, for $j=1,\ldots,S$.

\Cref{Thm2} below shows that, under the
appropriate alternative hypothesis, $LM_{1N}\left(
\hat{\bm\Lambda }\right)  $ and $LM_{2N}\left(
\hat{\bm\Lambda}\right)  $ diverge as sample sizes increase. Note that the condition $\sqrt{N}/T\rightarrow0$ is mild and is useful to ensure the consistency of the estimated factor loadings \citep{Bai2003}.

\begin{theorem}
\label{Thm2}Suppose that \Cref{Ass1,Ass2,Ass3} are satisfied. As
$N,T\rightarrow\infty,$ if $\sqrt{N}/T\rightarrow0,$ it holds that
\[
LM_{1N}\left(  \hat{\bm\Lambda}\right)
=N \delta_1+o_P(N),
\]
under the group heterogeneity alternative $H_{1}$, with $\delta_1$ a positive scalar, and%
\[
LM_{2N}\left(  \hat{\bm\Lambda}\right)
=N\delta_2+o_P(N),
\]
under the group heterogeneity alternative $H_{2}$, with $\delta_2$ a positive scalar.
\end{theorem}

The proof of \Cref{Thm2} is in Appendix A. In particular, we show that
$$\delta_1=\max_{1\leq j<k\leq S} \left(\left(\mathrm{Vech}\left(\bm{R}_{0}(j,k)\right)\right)^\prime\left(\bm{S}_0(j,k)\right)^{-1}\mathrm{Vech}\left(\bm{R}_{0}(j,k)\right)\right)$$
and
$$\delta_2=\min_{1\leq j<k\leq S}\left( \left(\mathrm{Vech}\left(\bm{R}_{0}(j,k)\right)\right)^\prime\left(\bm{S}_0(j,k)\right)^{-1}\mathrm{Vech}\left(\bm{R}_{0}(j,k)\right)\right).$$
Hence, under $H_1$, there is at least one group pair $(j,k)$ such that $\bm{R}_{0}(j,k) \neq \bm 0$, while under $H_2$, all group pairs
$(j,k)$ are such that $\bm{R}_{0}(j,k) \neq \bm 0$. Consequently, $\delta_1>0$ and $\delta_2>0$, under $H_1$ and under $H_2$, respectively.  Thus, $N\delta_1$ and  $N\delta_2$ diverge as $N$ goes to infinity,
and ensure that the test statistics $LM_{1N}\left(
\hat{\bm\Lambda }\right)  $ and $LM_{2N}\left(
\hat{\bm\Lambda}\right)  $, respectively, have power against the alternatives.
The decision to reject or not may be made using simulated
critical values from
$\max_{1\leq j<k\leq S}Q_{j,k}$
or
$\min_{1\leq j<k\leq S} Q_{j,k}$.
The following algorithm summarizes the proposed testing
procedure.

\begin{algorithm}
\label{Algo}\

\begin{enumerate}
\item Compute the estimated factor loadings ($\hat{\bm{\Lambda}}$)
: $\sqrt {N}$ times the eigenvectors corresponding to the $r$
largest eigenvalues of $\bm{XX}^{\prime} $,\ in decreasing order,
and using the normalization
$\hat{\bm{\Lambda}}^{\prime}\hat{\bm{\Lambda}}/N=\bm
I$.\footnote{Note that the results remain valid if the normalization $\hat{\bm
F}^{\prime} \hat{\bm F}/T=\bm I$ is used. However, in this case, the identity
matrix in the variance
formula should be replaced with $\hat{\bm \Lambda}^{\prime} \hat{\bm \Lambda}%
/N$.}

\item Find for $1\leq j<k\leq S,$%
\[
LM_{N}\left(  j,k,\hat{\bm{\Lambda}}\right)  =\bm{A}\left(
j,k,\hat {\bm{\Lambda}}\right)  ^{\prime}\left(
\hat{\bm{S}}\left(  j,k,\hat
{\bm{\Lambda}}\right)  \right)  ^{-1}\bm{A}\left(  j,k,\hat{\bm{\Lambda}}%
\right) ,
\]
where%
\[
\bm{A}\left(  j,k,\hat{\bm{\Lambda}}\right)
=\sqrt{N}\mathrm{Vech}\left(
\frac{1}{N_{j}}\sum_{i=M_{j-1}+1}^{M_{j}}\hat{\bm{\lambda}}_{i}\hat
{\bm{\lambda}}_{i}^{\prime}-\frac{1}{N_{k}}\sum_{i=M_{k-1}+1}^{M_{k}}%
\hat{\bm{\lambda}}_{i}\hat{\bm{\lambda}}_{i}^{\prime}\right) ,
\]
and%
\[
\bm{S}\left(  j,k,\hat{\bm{\Lambda}}\right)  =\left(
\frac{N}{N_{j}}+\frac {N}{N_{k}}\right)
\frac{1}{N}\sum_{i=1}^{N}\mathrm{Vech}\left(
\hat{\bm{\lambda}}_{i}\hat{\bm{\lambda
}}_{i}^{\prime}-\bm I\right)  \mathrm{Vech}\left(  \hat{\bm{\lambda}}_{i}%
\hat{\bm{\lambda}}_{i}^{\prime}-\bm I\right)  ^{\prime}.
\]

\item Obtain the test statistic%
\[
LM_{1N}\left(  \hat{\bm{\Lambda}}\right)  =\max_{1\leq j<k\leq
S}LM_{N}\left( j,k,\hat{\bm{\Lambda}}\right)
\]
or%
\[
LM_{2N}\left(  \hat{\bm{\Lambda}}\right)  =\min_{1\leq j<k\leq
S}LM_{N}\left( j,k,\hat{\bm{\Lambda}}\right)  .
\]

\item Reject $H_{0}$ or not using critical values
from
$\max_{1\leq j<k\leq S}Q_{j,k}$
or
$\min_{1\leq j<k\leq S} Q_{j,k}$.
\end{enumerate}
\end{algorithm}

Note that the number of factors is first selected using existing
selection criteria for factor models. To facilitate testings based on the proposed test statistics, we also propose a permutation approach in the next section.

\section{Permutation Inference\label{perm_results}}

The testing procedure described in \Cref{asym_results} relies on critical values obtained by simulating the
asymptotic distributions derived in \Cref{Thm1}.
In this section we consider an alternative inference procedure
that avoids the simulation of the asymptotic distributions,
and is based on a permutation approach that
approximates the cumulative distribution functions $F_{1}\left(  x\right)  $
and $F_{2}\left(  x\right)  $ of
$\max_{1\leq j<k\leq S}Q_{j,k}$
and
$\min_{1\leq j<k\leq S} Q_{j,k}$,
respectively. Permutation inference has been considered
in other contexts to test parametric or distributional hypotheses related to
observed variables. See, for example, \citet*{Fisher1936}, \citet*{LR2005} and
\citet*{HG2018}. An important feature of a permutation test is its ability
to have level $\alpha$ whenever the $S$\ populations of factor loadings have
distributions that are (approximately) the same under the null hypothesis.

Let $\bm G_{N}$ be the set of all possible permutations of the set $\left\{
1,\ldots,N\right\}  $ and $g\left(  i\right)  $ denote the value assigned by the
permutation $g$ to $i=1,\ldots,N.$
We also let
$g\hat{\bm{\Lambda}}=\left[  \hat{\lambda}_{g\left(  1\right)  }\cdots
\hat{\lambda}_{g\left(  N\right)  }\right]  ^{\prime}$ denote the matrix of factor
loadings based on the permutation $g.$

The permutation approach relies on \Cref{Ass4} (a) (Permutation Hypothesis)
stating that
the factor loadings $\left\{  \bm{\lambda}_{c,i}\right\}
_{i=1,\ldots,N}$ are identically distributed according to some distribution
$Q.$
\begin{assumption}
\label{Ass4}(Permutation Hypothesis)
\end{assumption}

\begin{description}
\item[(a)] The factor loadings $\left\{  \bm{\lambda}_{c,i}\right\}
_{i=1,\ldots,N}$ are identically distributed according to some distribution
$Q.$
\end{description}

This assumption, when combined with \Cref{Ass1} (b), ensures the exchangeability
hypothesis widely used in permutation contexts. Under this assumption, the
joint distribution of $\left(  \bm{\lambda}_{c,1},\ldots,\bm{\lambda}_{c,N}%
\right)  ^{\prime}$ is invariant under the permutations in $\bm{G}_{N}$. Therefore, for a
random permutation $g\in\bm{G}_{N},$ $\left(  \bm{\lambda}_{c,g\left(
1\right)  },\ldots,\bm{\lambda}_{c,g\left(  N\right)  }\right)  ^{\prime}$
has the same distribution as $\left(  \bm{\lambda}_{c,1},\ldots,\bm{\lambda}_{c,N}\right)
^{\prime}.$ The algorithm below summarizes the testing procedure based on the permutation approach.

\bigskip

\begin{algorithm1}
\label{Algo1}

\

All steps are identical to those in the algorithm in \Cref{Algo}, except for step 4,
which is replaced by the following:

\begin{enumerate}
\item[4.] A permutation approach is used to compute $P-$values\ by resampling
from $\left\{  \hat{\bm{\lambda}}_{1},\ldots,\hat{\bm{\lambda}}_{N}\right\}  $
without replacement to obtain $B$ random samples and computing the $P-$values:%
\[
P\text{-value}_{1}=\frac{1}{B+1}\left(  1+\sum_{b=1}^{B}\mathbb{I}\left(
LM_{1N}\left(  \hat{\bm{\Lambda}}\right)  \leq LM_{1N}\left(  g_{b}%
\hat{\bm{\Lambda}}\right)  \right)  \right)
\]
and%
\[
P\text{-value}_{2}=\frac{1}{B+1}\left(  1+\sum_{b=1}^{B}\mathbb{I}\left(
LM_{2N}\left(  \hat{\bm{\Lambda}}\right)  \leq LM_{2N}\left(  g_{b}%
\hat{\bm{\Lambda}}\right)  \right)  \right)  ,
\]
with $g_{b}\hat{\bm{\Lambda}}=\left[  \hat{\bm{\lambda}}_{g_{b}\left(
1\right)  }\ldots\hat{\bm{\lambda}}_{g_{b}\left(  N\right)  }\right]
^{\prime},$ $g_{b}\in\bm G_{N}$.
\end{enumerate}
\end{algorithm1}



We now show that the constructed procedure is asymptotical valid.
For $m=1,2,$\ consider the permutation distribution $R_{mN}\left(  x\right)  $
of $LM_{mN}\left(  \hat{\bm\Lambda}\right)  $ defined by%
\begin{equation}
R_{mN}\left(  x\right)  =\frac{1}{N!}\sum_{g\in\bm G_{N}}\mathbb{I}\left(
LM_{mN}\left(  g\hat{\bm{\Lambda}}\right)  \leq x\right)  . \label{Perm_dist}%
\end{equation}
We prove that, under the null, the
permutation distributions $R_{1N}\left(  x\right)  $ and $R_{2N}\left(
x\right)  $
asymptotically approximate the cumulative distribution functions
$F_{1}\left(  x\right)  $ and
$F_{2}\left(  x\right)$ of\\
$\max_{1\leq j<k\leq S}Q_{j,k}$
and
$\min_{1\leq j<k\leq S} Q_{j,k}$
respectively.

\begin{theorem}
\label{Thm3}Suppose that \Cref{Ass1,Ass2,Ass4} are satisfied. As
$N,T\rightarrow\infty,$ if $\sqrt{N}/T\rightarrow0,$ then under the null, the
permutation distribution of $LM_{mN}\left(  \hat{\bm{\Lambda}}\right)  ,$
given in \Cref{Perm_dist}, satisfies%
\begin{equation}
 R_{mN}\left(  x\right)  \overset{P}{\longrightarrow} F_{m}\left(  x\right), \label{Perm0}%
\end{equation}
for $m=1,2,$ and any real $x$.
\end{theorem}

The proof of \Cref{Thm3} is in Appendix B.
\Cref{Thm3} allows us to test for group-specific heterogeneity using all permutations in $\bm G_{N}$.
If all distinct permutations
of $\hat{\lambda}_{1},\ldots,\hat{\lambda}_{N}$ into $S$ groups of sizes
$N_{1},\ldots,N_{S}$ were drawn and the factor loadings were observed, then
the resulting test would be exact.
However, this may be computationally intensive
since we are in a high-dimensional
setting where $N$ goes to infinity and the number of possible permutations may be
very large.
For this reason, it is common practice to use a stochastic approximation of $R_{mN}\left(  x\right)$ defined by%
\[
R_{mN,B}\left(  x\right)  =\frac{1}{B+1}\left(  1+\sum_{b=1}^{B}%
\mathbb{I}\left(  LM_{mN}\left(  g_{b}\hat{\bm{\Lambda}}\right)  \leq
x\right)  \right)  ,m=1,2,
\]
where the permutations $g_{1},\ldots g_{B}$ are IID and
uniform over $\bm G_{N}$. See, for example, \citet*[Remark 2.2]{CRS2017}.
Similar approaches were used in different situations in \citet*{Romano1990} and
\citet*{SR2004}. 
%

In the next section,
we investigate the size and the power 
of the permutation tests by comparing their performance to the results from the asymptotic framework developed in  \Cref{asym_results}.

\section{Monte Carlo Illustrations \label{simulation}}

We use Monte Carlo simulations of six data-generating processes
(DGP) to evaluate the small-sample properties of the proposed
testing procedures described in Section 3 and Section 4. The DGPs are denoted
DGP 1-a, DGP 2-a, DGP 1-b, DGP 2-b, DGP 1-c, and DGP 2-c. DGP 1-a and DGP 2-a are
used to investigate the size of the test while the other DGPs help
evaluate the power of the test. In specifying the DGPs below,
$\mathrm{NID}$ means normally and identically distributed, and
$\mathrm{U}$ is the uniform distribution.

DGP 1-a assumes%
\begin{equation}
X_{it}=\lambda_{c,i}f_{c,t}+\kappa e_{it},\text{
}i=1,\ldots,N\text{ and }t=1,\ldots,T,
\end{equation}
with
\[
e_{it}\sim\mathrm{NID}\left(  0,1\right)  ,\text{ }f_{c,t}\sim\mathrm{NID}%
\left(  0,1\right)  ,\text{ }\lambda_{c,i}\sim\mathrm{NID}\left(
b,1\right) .
\]
We choose $\kappa=\sqrt{1+b^{2}}$, where $b$ is the mean of the
factor loadings, such that
$\mathrm{R}^{2}=1-\frac{\mathrm{trace}\left( \mathrm{E}\left(
\bm{ee}^{\prime}\right)  \right)  }{\mathrm{trace}\left(
\mathrm{E}\left(  \bm{XX}^{\prime}\right)  \right)  }=0.50$.
DGP 2-a allows cross-sectional dependence in
idiosyncratic
errors%
\begin{align}
e_{it}  &  =\sigma_{i}\left(  u_{it}+\sum_{1\leq\left\vert
j\right\vert \leq
P}\theta u_{\left(  i-j\right)  t}\right)  ,\text{ }u_{it}\sim\mathrm{NID}%
\left(  0,1\right)  ,\text{ }\label{idio}\\
\sigma_{i}  &  \sim U\left(  0.5,1.5\right)  \text{ and
}\kappa=\sqrt {\frac{12\left(  1+b^{2}\right)  }{13\left(
1+2P\theta^{2}\right)  }},
\end{align}
where $\theta=0.1$ and $P=4$.

In DGP 1-b and DGP 2-b, each group has a distinct specific factor.
We assume group-specific heterogeneity involving four groups and
one specific factor in each group, such that $g_{i}=1$ if
$i\in\left\{  1,\ldots,N/4\right\} ,g_{i}=2$ if $i\in\left\{
N/4+1,\ldots,N/2\right\}  ,$ $g_{i}=3$ if $i\in\left\{
N/2+1,\ldots,3N/4\right\}  $ and $g_{i}=4$ if $i\in\left\{
3N/4+1,\ldots,N\right\}  $, and%

\begin{equation}
X_{it}=\lambda_{c,i}f_{c,t}+\lambda_{g_{i},i}f_{g_{i},t}+\kappa
e_{it},\text{ }i=1,\ldots,N\text{ and }t=1,\ldots,T,
\end{equation}
with%
\[
\text{ }f_{g_{i},t}\sim\mathrm{NID}\left(  0,1\right)  ,\text{
}\lambda _{g_{i},i}\sim\mathrm{NID}\left(  b,1\right)  .
\]
We introduce a parameter $\rho$ representing the correlation
between any two specific factors and set $\rho=0.3.$
Similarly to DGP 2-a, DGP 2-b adds the cross-sectional dependence in idiosyncratic errors.

In all settings, we simulate the data $M=10,000$ times, set $b=1$
and use sample sizes $\left(  N,T\right)  $ that belong to
$\left\{ 80,120,160,200\right\}  \times\left\{  50,100\right\}  $.
We simulate the limit distribution for the asymptotic results. The
considered level is $5\%.$ The critical values are computed using
$500,000$ simulated data.
\bigskip

{\renewcommand{\arraystretch}{1.5}
{\setlength{\tabcolsep}{0.45cm}
\begin{table}[ptbh]
\caption{ Test rejection frequencies (\%) for DGP 1-a}%
\label{tab1a}
\centering%
\begin{tabular}
[c]{ccccc}\hline\hline
$LM_{1N}$ AsymptoticTest & \multicolumn{1}{|c}{$N=80$} &
\multicolumn{1}{|c|}{$N=120$} & \multicolumn{1}{|c}{$N=160$} &
\multicolumn{1}{|c}{$N=200$}\\\hline\hline
$T=50$ & \multicolumn{1}{|c}{4.35} & \multicolumn{1}{|c|}{4.47} &
\multicolumn{1}{|c}{4.54} & \multicolumn{1}{|c}{4.53}\\\hline
$T=100$ & \multicolumn{1}{|c}{4.13} & \multicolumn{1}{|c|}{4.32} &
\multicolumn{1}{|c}{4.41} & \multicolumn{1}{|c}{4.71}\\\hline\hline
&  &  &  & \\\hline\hline
$LM_{2N}$ Asymptotic Test & \multicolumn{1}{|c}{$N=80$} &
\multicolumn{1}{|c|}{$N=120$} & \multicolumn{1}{|c}{$N=160$} &
\multicolumn{1}{|c}{$N=200$}\\\hline\hline
$T=50$ & \multicolumn{1}{|c}{4.87} & \multicolumn{1}{|c|}{4.66} &
\multicolumn{1}{|c}{5.14} & \multicolumn{1}{|c}{4.72}\\\hline
$T=100$ & \multicolumn{1}{|c}{4.93} & \multicolumn{1}{|c|}{4.77} &
\multicolumn{1}{|c}{4.61} & \multicolumn{1}{|c}{5.04}\\\hline\hline
&  &  &  & \\\hline\hline
$LM_{1N}$ Permutation Test & \multicolumn{1}{|c}{$N=80$} &
\multicolumn{1}{|c|}{$N=120$} & \multicolumn{1}{|c}{$N=160$} &
\multicolumn{1}{|c}{$N=200$}\\\hline\hline
$T=50$ & \multicolumn{1}{|c}{4.86} & \multicolumn{1}{|c|}{4.50} &
\multicolumn{1}{|c}{4.73} & \multicolumn{1}{|c}{4.56}\\\hline
$T=100$ & \multicolumn{1}{|c}{4.60} & \multicolumn{1}{|c}{4.54} &
\multicolumn{1}{|c}{4.47} & \multicolumn{1}{|c}{4.65}\\\hline\hline
&  &  &  & \\\hline\hline
$LM_{2N}$ Permutation Test & \multicolumn{1}{|c}{$N=80$} &
\multicolumn{1}{|c|}{$N=120$} & \multicolumn{1}{|c}{$N=160$} &
\multicolumn{1}{|c}{$N=200$}\\\hline\hline
$T=50$ & \multicolumn{1}{|c}{4.86} & \multicolumn{1}{|c|}{4.60} &
\multicolumn{1}{|c}{4.99} & \multicolumn{1}{|c}{4.54}\\\hline
$T=100$ & \multicolumn{1}{|c}{5.01} & \multicolumn{1}{|c}{4.80} &
\multicolumn{1}{|c}{4.54} & \multicolumn{1}{|c}{4.91}\\\hline\hline
\end{tabular}
\par
Note: This table presents the rejection frequencies over $10,000$ simulated
datasets when there is no group specific factor and the level of the test is
$5\%$.\end{table}}}

{\renewcommand{\arraystretch}{1.5}
{\setlength{\tabcolsep}{0.45cm}
\begin{table}[ptbh]
\caption{ Test rejection frequencies (\%) for DGP 2-a}%
\label{tab2a}
\centering%
\begin{tabular}
[c]{ccccc}\hline\hline
$LM_{1N}$ Asymptotic Test & \multicolumn{1}{|c}{$N=80$} &
\multicolumn{1}{|c|}{$N=120$} & \multicolumn{1}{|c}{$N=160$} &
\multicolumn{1}{|c}{$N=200$}\\\hline\hline
$T=50$ & \multicolumn{1}{|c}{5.49} & \multicolumn{1}{|c|}{4.97} &
\multicolumn{1}{|c}{5.89} & \multicolumn{1}{|c}{5.64}\\\hline
$T=100$ & \multicolumn{1}{|c}{5.27} & \multicolumn{1}{|c|}{5.16} &
\multicolumn{1}{|c}{5.48} & \multicolumn{1}{|c}{5.42}\\\hline\hline
&  &  &  & \\\hline\hline
$LM_{2N}$ Asymptotic Test & \multicolumn{1}{|c}{$N=80$} &
\multicolumn{1}{|c|}{$N=120$} & \multicolumn{1}{|c}{$N=160$} &
\multicolumn{1}{|c}{$N=200$}\\\hline\hline
$T=50$ & \multicolumn{1}{|c}{5.03} & \multicolumn{1}{|c|}{5.55} &
\multicolumn{1}{|c}{5.43} & \multicolumn{1}{|c}{5.33}\\\hline
$T=100$ & \multicolumn{1}{|c}{5.20} & \multicolumn{1}{|c|}{5.33} &
\multicolumn{1}{|c}{5.48} & \multicolumn{1}{|c}{5.40}\\\hline\hline
&  &  &  & \\\hline\hline
$LM_{1N}$ Permutation Test & \multicolumn{1}{|c}{$N=80$} &
\multicolumn{1}{|c|}{$N=120$} & \multicolumn{1}{|c}{$N=160$} &
\multicolumn{1}{|c}{$N=200$}\\\hline\hline
$T=50$ & \multicolumn{1}{|c}{5.88} & \multicolumn{1}{|c|}{5.15} &
\multicolumn{1}{|c}{5.99} & \multicolumn{1}{|c}{5.82}\\\hline
$T=100$ & \multicolumn{1}{|c}{5.74} & \multicolumn{1}{|c}{5.42} &
\multicolumn{1}{|c}{5.54} & \multicolumn{1}{|c}{5.50}\\\hline\hline
&  &  &  & \\\hline\hline
$LM_{2N}$ Permutation Test & \multicolumn{1}{|c}{$N=80$} &
\multicolumn{1}{|c|}{$N=120$} & \multicolumn{1}{|c}{$N=160$} &
\multicolumn{1}{|c}{$N=200$}\\\hline\hline
$T=50$ & \multicolumn{1}{|c}{5.15} & \multicolumn{1}{|c|}{5.46} &
\multicolumn{1}{|c}{5.33} & \multicolumn{1}{|c}{5.07}\\\hline
$T=100$ & \multicolumn{1}{|c}{5.26} & \multicolumn{1}{|c}{5.22} &
\multicolumn{1}{|c}{5.28} & \multicolumn{1}{|c}{5.14}\\\hline\hline
\end{tabular}
\par
Note: See note for \Cref{tab1a}.\end{table}}}

{\renewcommand{\arraystretch}{1.5}
{\setlength{\tabcolsep}{0.45cm}
\begin{table}[ptbh]
\caption{ Test rejection frequencies (\%) for DGP 1-b}%
\label{tab1b}
\centering%
\begin{tabular}
[c]{ccccc}\hline\hline
$LM_{1N}$ Asymptotic Test & \multicolumn{1}{|c}{$N=80$} &
\multicolumn{1}{|c|}{$N=120$} & \multicolumn{1}{|c}{$N=160$} &
\multicolumn{1}{|c}{$N=200$}\\\hline\hline
$T=50$ & \multicolumn{1}{|c}{82.23} & \multicolumn{1}{|c|}{95.03} &
\multicolumn{1}{|c}{98.32} & \multicolumn{1}{|c}{99.40}\\\hline
$T=100$ & \multicolumn{1}{|c}{94.94} & \multicolumn{1}{|c|}{99.48} &
\multicolumn{1}{|c}{99.99} & \multicolumn{1}{|c}{100.00}\\\hline\hline
&  &  &  & \\\hline\hline
$LM_{2N}$ Asymptotic Test & \multicolumn{1}{|c}{$N=80$} &
\multicolumn{1}{|c|}{$N=120$} & \multicolumn{1}{|c}{$N=160$} &
\multicolumn{1}{|c}{$N=200$}\\\hline\hline
$T=50$ & \multicolumn{1}{|c}{50.50} & \multicolumn{1}{|c|}{75.09} &
\multicolumn{1}{|c}{86.68} & \multicolumn{1}{|c}{92.32}\\\hline
$T=100$ & \multicolumn{1}{|c}{76.95} & \multicolumn{1}{|c|}{94.00} &
\multicolumn{1}{|c}{99.10} & \multicolumn{1}{|c}{99.85}\\\hline\hline
&  &  &  & \\\hline\hline
$LM_{1N}$ Permutation Test & \multicolumn{1}{|c}{$N=80$} &
\multicolumn{1}{|c|}{$N=120$} & \multicolumn{1}{|c}{$N=160$} &
\multicolumn{1}{|c}{$N=200$}\\\hline\hline
$T=50$ & \multicolumn{1}{|c}{83.99} & \multicolumn{1}{|c|}{95.43} &
\multicolumn{1}{|c}{98.37} & \multicolumn{1}{|c}{99.46}\\\hline
$T=100$ & \multicolumn{1}{|c}{95.48} & \multicolumn{1}{|c}{99.48} &
\multicolumn{1}{|c}{99.99} & \multicolumn{1}{|c}{100.00}\\\hline\hline
&  &  &  & \\\hline\hline
$LM_{2N}$ Permutation Test & \multicolumn{1}{|c}{$N=80$} &
\multicolumn{1}{|c|}{$N=120$} & \multicolumn{1}{|c}{$N=160$} &
\multicolumn{1}{|c}{$N=200$}\\\hline\hline
$T=50$ & \multicolumn{1}{|c}{50.24} & \multicolumn{1}{|c|}{74.80} &
\multicolumn{1}{|c}{86.47} & \multicolumn{1}{|c}{92.14}\\\hline
$T=100$ & \multicolumn{1}{|c}{76.72} & \multicolumn{1}{|c}{93.91} &
\multicolumn{1}{|c}{99.08} & \multicolumn{1}{|c}{99.86}\\\hline\hline
\end{tabular}

\par
Note: This table presents the rejection frequencies over $10,000$ simulated
datasets when group specific factors arise and the level of the test is
$5\%$.\end{table}}}

{\renewcommand{\arraystretch}{1.5}
{\setlength{\tabcolsep}{0.45cm}
\begin{table}[ptbh]
\caption{ Test rejection frequencies (\%) for DGP 2-b}%
\label{tab2b}
\centering%
\begin{tabular}
[c]{ccccc}\hline\hline
$LM_{1N}$ Asymptotic Test & \multicolumn{1}{|c}{$N=80$} &
\multicolumn{1}{|c|}{$N=120$} & \multicolumn{1}{|c}{$N=160$} &
\multicolumn{1}{|c}{$N=200$}\\\hline\hline
$T=50$ & \multicolumn{1}{|c}{94.50} & \multicolumn{1}{|c|}{98.65} &
\multicolumn{1}{|c}{99.46} & \multicolumn{1}{|c}{99.80}\\\hline
$T=100$ & \multicolumn{1}{|c}{99.55} & \multicolumn{1}{|c|}{99.93} &
\multicolumn{1}{|c}{100.00} & \multicolumn{1}{|c}{100.00}\\\hline\hline
&  &  &  & \\\hline\hline
$LM_{2N}$ Asymptotic Test & \multicolumn{1}{|c}{$N=80$} &
\multicolumn{1}{|c|}{$N=120$} & \multicolumn{1}{|c}{$N=160$} &
\multicolumn{1}{|c}{$N=200$}\\\hline\hline
$T=50$ & \multicolumn{1}{|c}{76.77} & \multicolumn{1}{|c|}{89.29} &
\multicolumn{1}{|c}{93.99} & \multicolumn{1}{|c}{96.88}\\\hline
$T=100$ & \multicolumn{1}{|c}{94.43} & \multicolumn{1}{|c|}{98.74} &
\multicolumn{1}{|c}{99.81} & \multicolumn{1}{|c}{99.97}\\\hline\hline
&  &  &  & \\\hline\hline
$LM_{1N}$ Permutation Test & \multicolumn{1}{|c}{$N=80$} &
\multicolumn{1}{|c|}{$N=120$} & \multicolumn{1}{|c}{$N=160$} &
\multicolumn{1}{|c}{$N=200$}\\\hline\hline
$T=50$ & \multicolumn{1}{|c}{95.07} & \multicolumn{1}{|c|}{98.48} &
\multicolumn{1}{|c}{99.48} & \multicolumn{1}{|c}{99.80}\\\hline
$T=100$ & \multicolumn{1}{|c}{99.63} & \multicolumn{1}{|c}{99.93} &
\multicolumn{1}{|c}{100.00} & \multicolumn{1}{|c}{100.00}\\\hline\hline
&  &  &  & \\\hline\hline
$LM_{2N}$ Permutation Test & \multicolumn{1}{|c}{$N=80$} &
\multicolumn{1}{|c|}{$N=120$} & \multicolumn{1}{|c}{$N=160$} &
\multicolumn{1}{|c}{$N=200$}\\\hline\hline
$T=50$ & \multicolumn{1}{|c}{76.51} & \multicolumn{1}{|c|}{89.06} &
\multicolumn{1}{|c}{93.87} & \multicolumn{1}{|c}{96.82}\\\hline
$T=100$ & \multicolumn{1}{|c}{94.34} & \multicolumn{1}{|c}{98.76} &
\multicolumn{1}{|c}{99.80} & \multicolumn{1}{|c}{99.96}\\\hline\hline
\end{tabular}
\par
Note: See note for \Cref{tab1b}.\end{table}}}

When there is no group-specific heterogeneity, the $LM_{1N}$ and $LM_{2N}$ tests tend
to provide rejection frequencies close to the level of the test (see
\Cref{tab1a,tab2a}). When there is group-specific heterogeneity,
\Cref{tab1b,tab2b} show that the power of the tests generally increases with
the number of variables and the sample size. Also, the $LM_{1N}$ test is shown
to have a superior power compared to the $LM_{2N}$ test in all cases. However,
when the number of variables increases, the results show that the power of the
$LM_{2N}$ test increases rapidly and virtually catches up with the power of
the $LM_{1N}$ test. The findings are consistent with the fact that the second
test has a stronger alternative hypothesis than the first one. These results
also suggest that the testing procedures exhibit good control of size and
power. When the asymptotic distribution is approximated using the permutation
procedure, the results in \Cref{tab1a,tab2a,tab1b,tab2b} show that the size
is often around the $5\%$ level. 
Also, the power increases toward $100\%$
as the sample sizes increase, suggesting a good approximation of the
asymptotic distributions derived in \Cref{asym_results}.

{\renewcommand{\arraystretch}{1.5}
{\setlength{\tabcolsep}{0.45cm}
\begin{table}[ptbh]
\caption{ Test rejection frequencies (\%) for DGP 1-c}%
\label{tab1c}%
\centering%
\begin{tabular}
[c]{ccccc}\hline\hline
$LM_{1N}$ AsymptoticTest & \multicolumn{1}{|c}{$N=80$} &
\multicolumn{1}{|c|}{$N=120$} & \multicolumn{1}{|c}{$N=160$} &
\multicolumn{1}{|c}{$N=200$}\\\hline\hline
$T=50$ & \multicolumn{1}{|c}{81.27} & \multicolumn{1}{|c|}{94.96} &
\multicolumn{1}{|c}{98.58} & \multicolumn{1}{|c}{99.43}\\\hline
$T=100$ & \multicolumn{1}{|c}{93.42} & \multicolumn{1}{|c|}{99.55} &
\multicolumn{1}{|c}{99.55} & \multicolumn{1}{|c}{100.00}\\\hline\hline
&  &  &  & \\\hline\hline
{$LM_{2N}$ Asymptotic Test} & \multicolumn{1}{|c}{$N=80$} &
\multicolumn{1}{|c|}{$N=120$} & \multicolumn{1}{|c}{$N=160$} &
\multicolumn{1}{|c}{$N=200$}\\\hline\hline
{$T=50$} & \multicolumn{1}{|c}{14.54} & \multicolumn{1}{|c|}{11.50} &
\multicolumn{1}{|c}{11.31} & \multicolumn{1}{|c}{9.58}\\\hline
{$T=100$} & \multicolumn{1}{|c}{10.84} & \multicolumn{1}{|c|}{8.75} &
\multicolumn{1}{|c}{7.40} & \multicolumn{1}{|c}{7.25}\\\hline\hline
&  &  &  & \\\hline\hline
{$LM_{1N}$ Permutation Test} & \multicolumn{1}{|c}{$N=80$} &
\multicolumn{1}{|c|}{$N=120$} & \multicolumn{1}{|c}{$N=160$} &
\multicolumn{1}{|c}{$N=200$}\\\hline\hline
{$T=50$} & \multicolumn{1}{|c}{83.96} & \multicolumn{1}{|c|}{95.52} &
\multicolumn{1}{|c}{98.76} & \multicolumn{1}{|c}{99.44}\\
{$T=100$} & \multicolumn{1}{|c}{94.70} & \multicolumn{1}{|c}{99.63} &
\multicolumn{1}{|c}{99.98} & \multicolumn{1}{|c}{100.00}\\\hline\hline
&  &  &  & \\\hline\hline
{$LM_{2N}$ Permutation Test} & \multicolumn{1}{|c}{$N=80$} &
\multicolumn{1}{|c|}{$N=120$} & \multicolumn{1}{|c}{$N=160$} &
\multicolumn{1}{|c}{$N=200$}\\\hline\hline
{$T=50$} & \multicolumn{1}{|c}{14.01} & \multicolumn{1}{|c|}{11.04} &
\multicolumn{1}{|c}{10.91} & \multicolumn{1}{|c}{9.33}\\\hline
{$T=100$} & \multicolumn{1}{|c}{10.29} & \multicolumn{1}{|c}{8.30} &
\multicolumn{1}{|c}{7.06} & \multicolumn{1}{|c}{6.95}\\\hline\hline
\end{tabular}
\par
Note: See note for \Cref{tab1b}.\end{table}}}

{\renewcommand{\arraystretch}{1.5}
{\setlength{\tabcolsep}{0.45cm}
\begin{table}[ptbh]
\caption{ Test rejection frequencies (\%) for DGP 2-c}%
\label{tab2c}%
\centering%
\begin{tabular}
[c]{ccccc}\hline\hline
{$LM_{1N}$ Asymptotic Test} & \multicolumn{1}{|c}{$N=80$} &
\multicolumn{1}{|c|}{$N=120$} & \multicolumn{1}{|c}{$N=160$} &
\multicolumn{1}{|c}{$N=200$}\\\hline\hline
{$T=50$} & \multicolumn{1}{|c}{91.06} & \multicolumn{1}{|c|}{97.69} &
\multicolumn{1}{|c}{99.15} & \multicolumn{1}{|c}{99.60}\\\hline
{$T=100$} & \multicolumn{1}{|c}{97.64} & \multicolumn{1}{|c|}{99.84} &
\multicolumn{1}{|c}{99.99} & \multicolumn{1}{|c}{100.00}\\\hline\hline
&  &  &  & \\\hline\hline
{$LM_{2N}$ Asymptotic Test} & \multicolumn{1}{|c}{$N=80$} &
\multicolumn{1}{|c|}{$N=120$} & \multicolumn{1}{|c}{$N=160$} &
\multicolumn{1}{|c}{$N=200$}\\\hline\hline
{$T=50$} & \multicolumn{1}{|c}{23.08} & \multicolumn{1}{|c|}{20.47} &
\multicolumn{1}{|c}{19.85} & \multicolumn{1}{|c}{18.32}\\\hline
{$T=100$} & \multicolumn{1}{|c}{20.55} & \multicolumn{1}{|c|}{17.41} &
\multicolumn{1}{|c}{16.29} & \multicolumn{1}{|c}{16.06}\\\hline\hline
&  &  &  & \\\hline\hline
{$LM_{1N}$ Permutation Test} & \multicolumn{1}{|c}{$N=80$} &
\multicolumn{1}{|c|}{$N=120$} & \multicolumn{1}{|c}{$N=160$} &
\multicolumn{1}{|c}{$N=200$}\\\hline\hline
{$T=50$} & \multicolumn{1}{|c}{92.44} & \multicolumn{1}{|c|}{97.92} &
\multicolumn{1}{|c}{99.30} & \multicolumn{1}{|c}{99.59}\\\hline
{$T=100$} & \multicolumn{1}{|c}{98.16} & \multicolumn{1}{|c}{99.86} &
\multicolumn{1}{|c}{99.99} & \multicolumn{1}{|c}{100.00}\\\hline\hline
&  &  &  & \\\hline\hline
{$LM_{2N}$ Permutation Test} & \multicolumn{1}{|c}{$N=80$} &
\multicolumn{1}{|c|}{$N=120$} & \multicolumn{1}{|c}{$N=160$} &
\multicolumn{1}{|c}{$N=200$}\\\hline\hline
{$T=50$} & \multicolumn{1}{|c}{22.04} & \multicolumn{1}{|c|}{19.94} &
\multicolumn{1}{|c}{19.41} & \multicolumn{1}{|c}{17.67}\\\hline
{$T=100$} & \multicolumn{1}{|c}{19.53} & \multicolumn{1}{|c}{16.51} &
\multicolumn{1}{|c}{15.53} & \multicolumn{1}{|c}{15.0}\\\hline\hline
\end{tabular}
\par
Note: See note for \Cref{tab1b}.\end{table}}}

To illustrate a situation where the $LM_{1N}$ and $LM_{2N}$ tests
may provide different outcomes, we consider
the designs DGP 1-c and DGP 2-c. They are constructed by setting $f_{1,t}%
=f_{2,t}=f_{3,t}$, in DGP 1-b and DGP 2-b, respectively, so that
the first three groups of variables are affected by the same
specific factor.
Similarly to DGP 2-a, DGP 2-c adds the cross-sectional dependence in idiosyncratic errors.
As expected, the $LM_{2N}$ test has very low rejection
frequencies, while the $LM_{1N}$ test shows very high rejection
frequencies (see \Cref{tab1c,tab2c}). Indeed, by construction, the
$LM_{1N}$ should be used if the purpose is to test whether
group-specific heterogeneity appears in at least one pair of groups of
variables and the $LM_{2N}$ is useful for testing if group-specific heterogeneity appears in all pairs of groups. 

\section{Empirical Illustration}
\label{application}

This section illustrates the proposed statistical tests with an
empirical application. The purpose of this application is to study the existence of
heterogeneous groups of industries within the U.S. financial sector. It
contributes to the understanding of industry groups and is
related to other financial applications that analyze the
performance of stocks relative to their membership in industry
groups, or the propagation of systematic risks to a whole industry
and the overall economy. See, for example, \citet*{KW1996},
\citet*{NB2005}, and \citet*{HS2013}, for the role played by
industry groups in earnings management, event studies, and risk
management.

Originally created in 1937 in the United States, the Standard
Industrial Classification (SIC) was designed to identify
homogenous groups of companies based on the nature of the
production process and product characteristics.
SIC codes for U.S. stocks are available from the Center for
Research in Security Prices (CRSP). To conduct our empirical
analysis, we use data on daily stock prices from the CRSP U.S.
stock database, for the companies in the Finance, Insurance, and
Real Estate division of the SIC. The sample consists of daily log
returns for the period from January 3, 2018, to December 31, 2019,
and includes 502 observations for each stock. Stocks with
insufficient or missing observations are discarded.

The SIC system classifies companies into industries, industries
into industry groups, industry groups into major groups, and major
groups into divisions. We consider the classifications according
to major groups and industry groups and, in both cases, keep only
the groups for which at least 30 stocks are available in the
sample. The definitions of the resulting groups and the
corresponding number of stocks are shown in \Cref{tab7} and  \Cref{tab8}. The
number of stocks in the groups ranges from 30 to 2331.

{\renewcommand{\arraystretch}{1.5}
{\setlength{\tabcolsep}{0.45cm}
\begin{table}[ptbh]
\caption{ The SIC major groups in the application}%
\label{tab7}
\centering%
\begin{tabular}
[c]{l}\hline\hline
SIC Name (Number of stocks in the sample)\\\cline{1-1}%
Depository institutions (307)\\
Nondepository credit institutions (36)\\
Security and commodity brokers, dealers, exchanges, and services (101)\\
Insurance carriers (94)\\
Real estate (45)\\
Holding and other investment offices (2331)\\\hline\hline
\end{tabular}
\end{table}}}

{\renewcommand{\arraystretch}{1.5}
{\setlength{\tabcolsep}{0.45cm}
\begin{table}[ptbh]
\caption{ The SIC industry groups in the application}%
\label{tab8}
\centering%
\begin{tabular}
[c]{l}\hline\hline
SIC Name (Number of stocks in the sample)\\\cline{1-1}%
Commercial banks (237)\\
Savings institutions (67)\\
Security brokers, dealers, and flotation companies (30)\\
Services allied with the exchange of securities or commodities (51)\\
Fire, marine, and casualty insurance (39)\\
Holding offices (36)\\
Investment offices (2045)\\
Trusts (48)\\
Miscellaneous investing (201)\\\hline\hline
\end{tabular}
\end{table}}}

Using these data, we answer the following question. Can we find
statistical evidence of 
heterogeneous SIC major
groups and SIC industry groups?
To answer this question, we apply the two statistical
tests to check for evidence of group-specific heterogeneity. The
results for the asymptotic tests are summarized in  \Cref{tab9}, which also includes the 
simulated critical values. For the current application, we have $S=6$ and $r=10$ for the major groups
classification, and $S=9$ and $r=12$ for the industry groups classification. 

\Cref{tab9} shows that, for both group classifications, the $LM_{1N}$
test clearly rejects the null hypothesis of no group-specific
heterogeneity. The $LM_{2N}$ test rejects the null hypothesis for
the major groups, but not for the industry groups. These
results suggest the following: 
there is at least one pair of heterogeneous industry groups, and all pairs of major industry groups are heterogeneous.
When the permutation approach is
used, we obtain the same results. In particular, for the major groups
classification, the $P$ values for the $LM_{1N}$ and $LM_{2N}$ tests are both
$0.001$, while for the industry groups classification, the $P$ values
for the two tests
are $0.001$ and $0.991$,
respectively.

{\renewcommand{\arraystretch}{1.5}
{\setlength{\tabcolsep}{0.45cm}
\begin{table}[ptbh]
\caption{ The results of the statistical tests}%
\label{tab9}
\centering%
\begin{tabular}
[c]{l|c|c}\hline\hline
& SIC major groups & SIC industry groups\\\cline{1-3}%
$LM_{1N}$ Test & 1492.64 & 1380.83\\
Critical value & 87.76 & 119.11\\\cline{1-3}%
$LM_{2N}$ Test & 87.94 & 38.67\\
Critical value & 47.41 & 65.44\\\hline\hline
\end{tabular}
\end{table}}}


\section{Conclusion \label{conclusion}}

This article contributes to the literature on factor models by developing two tests for group-specific heterogeneity.
Both tests have the same null hypothesis under which there is no group-specific heterogeneity.
They differ in the type of alternative hypothesis considered. For the first test, it states that group-specific heterogeneity emerges in at least one pair of groups; for the second test, it states that group-specific heterogeneity emerges in all pairs of groups.
We derive the asymptotic distributions of the test statistics
under the null hypothesis, and prove that each test statistics diverges under
the corresponding alternative hypothesis.

In addition, we propose and show the validity of permutation tests that can be used
as an alternative to the asymptotic tests, if an exchangeability condition on the factor loadings is satisfied.
Interestingly, given the normalization of the proposed test statistics, arguments similar to those in \citet*{CR2013} can be used to allow more heterogeneous factor loadings at the cost of more cumbersome derivations. However, this point is beyond the scope of this paper and is left for future research.

The proposed tests have the advantage that the estimation of the factors and factor
loadings under the alternatives is not required. Results based on several
simulation experiments are presented to illustrate the performance
of the tests. An empirical investigation that detects the existence of heterogeneous 
groups of industries within the U.S. financial sector shows the empirical
relevance of the proposed tests.




\section*{Appendix: Proofs of Results in \Cref{asym_results,perm_results} \label{appendix}}

\subsection*{Appendix A: Proofs of Results in Section 3 \label{appendix_2}}

\renewcommand{\theequation}{A.\arabic{equation}} \renewcommand{\thesection}{A} \setcounter{equation}{0}

This appendix proves the validity of the suggested test statistics under the
null and under the different alternatives. To prove \Cref{Thm1} an
\Cref{Thm2}, we rely on the following auxiliary results.

\begin{lemma}
\label{A1}Suppose that \Cref{Ass1,Ass2} are satisfied. If as $N,T\rightarrow
\infty,$ $\sqrt{N}/T\rightarrow0$, then under the null, it holds that
\begin{equation}
\frac{1}{N_{j}}\sum_{i=M_{j-1}+1}^{M_{j}}\left(  \hat{\bm{\lambda}}_{i}%
-\hat{\bm{H}}^{-1}\bm{\lambda}_{c,i}\right)  \bm{\lambda}_{i}^{\prime}%
=O_{P}\left(  \frac{1}{\delta_{NT}^{2}}\right)  , \label{A11}%
\end{equation}%
\begin{equation}
\frac{1}{N_{j}}\sum_{i=M_{j-1}+1}^{M_{j}}\left\Vert \hat{\bm{\lambda}}%
_{i}-\hat{\bm{H}}^{-1}\bm{\lambda}_{c,i}\right\Vert ^{2}=O_{P}\left(  \frac
{1}{\delta_{NT}^{2}}\right)  , \label{A13}%
\end{equation}
for any $j=1,\ldots,S.$
\end{lemma}

\begin{lemma}
\label{low_level} Suppose that \Cref{Ass1} (b), (g) and (h) are satisfied. As
$N\rightarrow\infty$, under the null, it holds that%
\begin{equation}
\mathop{\rm plim}_{N\rightarrow\infty}\bm{S}\left(  j,k,\bm{\Lambda}
\bm{H}_{0}^{\prime-1}\right)  =\lim_{N\rightarrow\infty}\mathrm{Var}\left(
\bm{A}\left(  j,k,\bm{\Lambda}\bm{H}_{0}^{\prime-1}\right)  \right)  ,
\label{low_level1}%
\end{equation}
for any $j$ and $k$ such that $1\leq j<k\leq S,$ and%
\begin{equation}
\bm{A}_{0}\left(  j,\bm{\Lambda}\bm{H}_{0}^{\prime-1}\right)
\overset{d}{\longrightarrow}\mathrm{N}\left(  \bm{0},\lim_{N\rightarrow\infty
}\mathrm{Var}\left(  \bm{A}_{0}\left(  j,\bm{\Lambda}\bm{H}_{0}^{\prime
-1}\right)  \right)  \right)  , \label{low_level2}%
\end{equation}
for any $j$ such that $1\leq j\leq S.$
\end{lemma}

\begin{lemma}
\label{A0}Suppose that \Cref{Ass1,Ass2} are satisfied. As $N,T\rightarrow
\infty,$ if $\sqrt{N}/T\rightarrow0$, then for any $j$ and $k$ such that
$1\leq j<k\leq S,$ under the null, it holds that,%

\begin{equation}
\left\Vert \bm{A}\left(  j,k,\hat{\bm{\Lambda}}\right)  -\bm{A}\left(
j,k,\bm{\Lambda}\bm{H}_{0}^{\prime-1}\right)  \right\Vert =o_{P}\left(
1\right)  \label{A01}%
\end{equation}
and%
\begin{equation}
\left\Vert \bm{S}\left(  j,k,\hat{\bm{\Lambda}}\right)  -\bm{S}\left(
j,k,\bm{\Lambda}\bm{H}_{0}^{\prime-1}\right)  \right\Vert =o_{P}\left(
1\right)  . \label{A02}%
\end{equation}

\end{lemma}

\begin{lemma}
\label{A2} Suppose that \Cref{Ass1,Ass2,Ass3} are satisfied. If as
$N,T\rightarrow\infty,$ $\sqrt{N}/T\rightarrow0$, then for any positive
integer $j$ such that $1\leq j\leq S,$ under the alternatives, it holds that
\begin{equation}
\frac{1}{N_{j}}\sum_{i=M_{j-1}+1}^{M_{j}}\left(  \hat{\bm{\lambda}}_{i}%
-\hat{\bm{\Xi}}^{-1}\bm{\phi}_{i}\right)  \bm{\phi}_{i}^{\prime}=O_{P}\left(
\delta_{NT}^{-2}\right)  , \label{A21}%
\end{equation}%
\begin{equation}
\frac{1}{N_{j}}\sum_{i=M_{j-1}+1}^{M_{j}}\left\Vert \hat{\bm{\lambda}}%
_{i}-\hat{\bm{\Xi}}^{-1}\bm{\phi}_{i}\right\Vert ^{2}=O_{P}\left(  \delta
_{NT}^{-2}\right)  , \label{A22}%
\end{equation}
and%
\begin{equation}
\left\Vert \bm{S}\left(  j,k,\hat{\bm{\Lambda}}\right)  -\bm{S}\left(
j,k,\bm{\Phi}\bm{\Xi}_{0}^{\prime-1}\right)  \right\Vert =o_{P}\left(
1\right)  . \label{A25}%
\end{equation}

\end{lemma}

Since \Cref{A1}, \Cref{low_level} \Cref{low_level1} , \Cref{A0} and \Cref{A2}
can be proved following steps nearly identical to those in \citet*[Lemmas A1,
A2, 2.1 and A3]{Djogbenou2019}, they are omitted. Moreover, the proof for
\Cref{low_level} \Cref{low_level2}, which is similar to the proof that (A.21)
in \citet*{Djogbenou2019} is asymptotically normal, is also omitted. We next
present the proofs of \Cref{Thm1} and \Cref{Thm2}.

\noindent\textbf{Proof of \Cref{Thm1}}

The proof consists of three parts. In the first and the second parts, we show
that
\[
\max_{1\leq j<k\leq S}\left\vert LM_{N}\left(  j,k,\hat{\bm{\Lambda}}\right)
-LM_{N}\left(  j,k,\bm{\Lambda}\bm{H}_{0}^{\prime-1}\right)  \right\vert
=o_{P}\left(  1\right)
\]
and%
\[
\max_{1\leq j<k\leq S}\left\vert LM_{N}\left(  j,k,\bm{\Lambda} {\bm{H}}_{0}%
^{\prime-1}\right)  -LM_{0N}\left(  j,k,\bm{\Lambda} {\bm{H}}_{0}^{\prime
-1}\right)  \right\vert =o_{P}\left(  1\right)  ,
\]
respectively, where%
\[
LM_{0N}\left(  j,k,\bm{\Lambda}\bm{H}_{0}^{\prime-1}\right)  =\bm{A}\left(
j,k,\bm{\Lambda}\bm{H}_{0}^{\prime-1}\right)  ^{\prime}\left(  \left(
\frac{1}{\pi_{j}}+\frac{1}{\pi_{k}}\right)  \bm{S}_{0}\right)  ^{-1}%
\bm{A}\left(  j,k,\bm{\Lambda}\bm{H}_{0}^{\prime-1}\right)  .
\]
In the third part, we derive the asymptotic distribution in \eqref{Asym1} and
\eqref{Asym2} after using the previous result to establish that $LM_{1N}%
\left(  \hat{\bm{\Lambda }}\right)  $ and $LM_{2N}\left(  \hat{\bm{\Lambda }}%
\right)  $ have the same asymptotic distributions as $\max_{1\leq j<k\leq
S}LM_{0N}\left(  j,k,\bm{\Lambda}\bm{H}_{0}^{\prime-1}\right)  $ and
$\min_{1\leq j<k\leq S}LM_{0N}\left(  j,k,\bm{\Lambda}\bm{H}_{0}^{\prime
-1}\right)  ,$ respectively.

\textbf{Part 1:} Using the triangle inequality, we have%
\[
\max_{1\leq j<k\leq S}\left\vert LM_{N}\left(  j,k,\hat{\bm{\Lambda}}\right)
-LM_{N}\left(  j,k,\bm{\Lambda}\bm{H}_{0}^{\prime-1}\right)  \right\vert \leq
M_{1}+M_{2}+M_{3},
\]
where%
\begin{align*}
M_{1}  &  =\max_{1\leq j<k\leq S}\left(  \left\Vert \bm{A}\left(
j,k,\hat{\bm{\Lambda}}\right)  \right\Vert \left\Vert \bm{S}\left(
j,k,\hat{\bm{\Lambda}}\right)  ^{-1}-\bm{S}\left(
j,k,\bm{\Lambda} {\bm{H}}_{0}^{\prime-1}\right)  ^{-1}\right\Vert \left\Vert
\bm{A}\left(  j,k,\hat{\bm{\Lambda}}\right)  \right\Vert \right)  ,\\
M_{2}  &  =\max_{1\leq j<k\leq S}\left(  \left\Vert \bm{A}\left(
j,k,\hat{\bm{\Lambda}}\right)  -\bm{A}\left(  j,k,\bm{\Lambda} {\bm{H}}_{0}%
^{\prime-1}\right)  \right\Vert \left\Vert \bm{S}\left(
j,k,\bm{\Lambda}\bm{H}_{0}^{\prime-1}\right)  ^{-1}\right\Vert \left\Vert
\bm{A}\left(  j,k,\hat{\bm{\Lambda }}\right)  \right\Vert \right)
\end{align*}
and%
\[
M_{3}=\max_{1\leq j<k\leq S}\left(  \left\Vert \bm{A}\left(
j,k,\bm{\Lambda}\bm{H}_{0}^{\prime-1}\right)  \right\Vert \left\Vert
\bm{S}\left(  j,k,\bm{\Lambda}\bm{H}_{0}^{\prime-1}\right)  ^{-1}\right\Vert
\left\Vert \bm{A}\left(  j,k,\hat{\bm{\Lambda}}\right)  -\bm{A}\left(
j,k,\bm{\Lambda}\bm{H}_{0}^{\prime-1}\right)  \right\Vert \right)  .
\]
The proof uses the following auxiliary results that hold for any $j$ and $k$
such that $1\leq j<k\leq S:$ $\left(  a\right)  $ $\left\Vert \bm{A}\left(
j,k,\bm{\Lambda}\bm{H}_{0}^{\prime-1}\right)  \right\Vert =O_{P}\left(
1\right)  ,\left(  b\right)  $ $\left\Vert \bm{A}\left(  j,k,\hat
{\bm{\Lambda}}\right)  -\bm{A}\left(  j,k,\bm{\Lambda}\bm{H}_{0}^{\prime
-1}\right)  \right\Vert =o_{P}\left(  1\right)  ,$ and $\left\Vert
\bm{A}\left(  j,k,\hat{\bm{\Lambda}}\right)  \right\Vert =O_{P}\left(
1\right)  ,$ and finally $\left(  c\right)  $ $\left\Vert \bm{S}\left(
j,k,\hat{\bm{\Lambda}}\right)  ^{-1}-\bm{S}\left(
j,k,\bm{\Lambda} {\bm{H}}_{0}^{\prime-1}\right)  ^{-1}\right\Vert
=o_{P}\left(  1\right)  $ and $\left\Vert \bm{S}\left(
j,k,\bm{\Lambda}\bm{H}_{0}^{\prime-1}\right)  ^{-1}\right\Vert =O_{P}\left(
1\right)  .$ Result $\left(  a\right)  $\ follows from \Cref{low_level}
\eqref{low_level2} and $\bm{A}\left(  j,k,\bm{\Lambda}\bm{H}_{0}^{\prime-1}\right)
=\bm{A}_{0}\left(  j,\bm{\Lambda}\bm{H}_{0}^{\prime-1}\right)  -\bm{A}_{0}\left(
k,\bm{\Lambda}\bm{H}_{0}^{\prime-1}\right)  .$ Result $\left(  b\right)  $
follows from \Cref{A0} \eqref{A01} and%
\[
\left\Vert \bm{A}\left(  j,k,\hat{\bm{\Lambda}}\right)  \right\Vert
\leq\left\Vert \bm{A}\left(  j,k,\hat{\bm{\Lambda
}}\right)  - \bm{A}\left(  j,k,\bm{\Lambda}\bm{H}_{0}^{\prime-1}\right)  \right\Vert
+\left\Vert \bm{A}\left(  j,k,\bm{\Lambda}\bm{H}_{0}^{\prime-1}\right)
\right\Vert =o_{P}\left(  1\right)  +O_{P}\left(  1\right)  =O_{P}\left(
1\right)  ,
\]
for any $j$ and $k.$ Result $\left(  c\right)  $ holds since for any $j$ and
$k,$ $\left\Vert \bm{S}\left(  j,k,\hat{\bm{\Lambda}}\right)  ^{-1}%
-\bm{S}\left(  j,k,\bm{\Lambda}\bm{H}_{0}^{\prime-1}\right)  ^{-1}\right\Vert
$ is bounded by
\begin{align*}
&  \left\Vert \bm{S}\left(  j,k,\bm{\Lambda}\bm{H}_{0}^{\prime-1}\right)
^{-1}\left(  \bm{S}\left(  j,k,\bm{\Lambda}\bm{H}_{0}^{\prime-1}\right)
-\bm{S}\left(  j,k,\hat{\bm{\Lambda}}\right)  \right)  \bm{S}\left(
j,k,\hat{\bm{\Lambda}}\right)  ^{-1}\right\Vert \\
&  \leq\left\Vert \bm{S}\left(  j,k,\bm{\Lambda}\bm{H}_{0}^{\prime-1}\right)
^{-1}\right\Vert \left\Vert \bm{S}\left(  j,k,\hat{\bm{\Lambda}}\right)
-\bm{S}\left(  j,k,\bm{\Lambda}\bm{H}_{0}^{\prime-1}\right)  \right\Vert
\left\Vert \bm{S}\left(  j,k,\hat{\bm{\Lambda}}\right)  ^{-1}\right\Vert ,
\end{align*}
$\left\Vert \bm{S}\left(  j,k,\hat{\bm{\Lambda}}\right)  -\bm{S}\left(
j,k,\bm{\Lambda}\bm{H}_{0}^{\prime-1}\right)  \right\Vert
\overset{P}{\longrightarrow}\bm{0}$ by \Cref{A0} \eqref{A02}, and
$\bm{S}\left(  j,k,\bm{\Lambda}\bm{H}_{0}^{\prime-1}\right)  ^{-1}$\ and
$\bm{S}\left(  j,k,\hat{\bm{\Lambda}}\right)  ^{-1}$\ are bounded in
probability by \Cref{low_level} \eqref{low_level1} and \Cref{A0} \eqref{A02},
and \Cref{Ass1} (g). From (a), (b) and (c), $\max_{1\leq j<k\leq S}\left\vert
LM_{N}\left(  j,k,\hat{\bm{\Lambda}}\right)  -LM_{N}\left(  j,k,\bm{\Lambda
}\bm{H}_{0}^{\prime-1}\right)  \right\vert =o_{P}\left(  1\right)  .$

\textbf{Part 2:} Noting that for any $j$ and $k$ such that $1\leq j<k\leq S,$%
\[
\left\vert LM_{N}\left(  j,k,\bm{\Lambda}\bm{H}_{0}^{\prime-1}\right)
-LM_{0N}\left(  j,k,\bm{\Lambda}\bm{H}_{0}^{\prime-1}\right)  \right\vert \leq
J_{1}+J_{2},
\]
where $J_{1}$ equals
\[
\max_{1\leq j<k\leq S}\left(  \left\Vert \bm{A}\left(
j,k,\bm{\Lambda}\bm{H}_{0}^{\prime-1}\right)  \right\Vert \left\Vert \left(
\mathrm{Var}\left(  \bm{A}\left(  j,k,\bm{\Lambda} {\bm{H}}_{0}^{\prime
-1}\right)  \right)  \right)  ^{-1}-\left(  \left(  \frac{1}{\pi_{j}}+\frac
{1}{\pi_{k}}\right)  \bm{S}_{0}\right)  ^{-1}\right\Vert \left\Vert
\bm{A}\left(  j,k,\bm{\Lambda}\bm{H}_{0}^{\prime-1}\right)  \right\Vert
\right)
\]
and $J_{2}$ equals
\[
\max_{1\leq j<k\leq S}\left(  \left\Vert \bm{A}\left(
j,k,\bm{\Lambda}\bm{H}_{0}^{\prime-1}\right)  \right\Vert \left\Vert \left(
\bm{S}\left(  j,k,\bm{\Lambda}\bm{H}_{0}^{\prime-1}\right)  \right)
^{-1}-\left(  \mathrm{Var}\left(  \bm{A}\left(  j,k,\bm{\Lambda}\bm{H}_{0}%
^{\prime-1}\right)  \right)  \right)  ^{-1}\right\Vert \left\Vert
\bm{A}\left(  j,k,\bm{\Lambda} {\bm{H}}_{0}^{\prime-1}\right)  \right\Vert
\right)  .
\]
Because $\max_{1\leq j<k\leq S}\left\Vert \bm{A}\left(
j,k,\bm{\Lambda} {\bm{H}}_{0}^{\prime-1}\right)  \right\Vert $ is
$O_{P}\left(  1\right)  $ from Part 1, we only need to show that
\begin{equation}
\max_{1\leq j<k\leq S}\left\Vert \left(  \mathrm{Var}\left(  \bm{A}\left(
j,k,\bm{\Lambda}\bm{H}_{0}^{\prime-1}\right)  \right)  \right)  ^{-1}-\left(
\left(  \frac{1}{\pi_{j}}+\frac{1}{\pi_{k}}\right)  \bm{S}_{0}\right)
^{-1}\right\Vert =o_{P}\left(  1\right)  \label{J1}%
\end{equation}
and%
\begin{equation}
\max_{1\leq j<k\leq S}\left\Vert \left(  \bm{S}\left(
j,k,\bm{\Lambda} {\bm{H}}_{0}^{\prime-1}\right)  \right)  ^{-1}-\left(
\mathrm{Var}\left(  \bm{A}\left(  j,k,\bm{\Lambda}\bm{H}_{0}^{\prime
-1}\right)  \right)  \right)  ^{-1}\right\Vert =o_{P}\left(  1\right)  .
\label{J2}%
\end{equation}

Using \Cref{Ass1} (g), we find that%
\[
\lim_{N\rightarrow\infty}\mathrm{Var}\left(  \bm{A}_{0}\left(
j,\bm{\Lambda}\bm{H}_{0}^{\prime-1}\right)  \right)  =\frac{1}{\pi_{j}%
}\bm{S}_{0},
\]
and we deduce that%
\[
\mathrm{Var}\left(  \bm{A}\left(  j,k,\bm{\Lambda}\bm{H}_{0}^{\prime
-1}\right)  \right)  =\mathrm{Var}\left(  \bm{A}_{0}\left(
j,\bm{\Lambda}\bm{H}_{0}^{\prime-1}\right)  \right)  +\mathrm{Var}\left(
\bm{A}_{0}\left(  k,\bm{\Lambda}\bm{H}_{0}^{\prime-1}\right)  \right)
\longrightarrow\left(  \frac{1}{\pi_{j}}+\frac{1}{\pi_{k}}\right)
\bm{S}_{0},
\]
where the limit is positive definite. Therefore, \Cref{J1} holds. Furthermore,
\Cref{low_level} \eqref{low_level1} implies $\bm{S}\left(
j,k,\bm{\Lambda}\bm{H}_{0}^{\prime-1}\right)  $ converges in probability to
the limit of $\mathrm{Var}\left(  \bm{A}\left(  j,k,\bm{\Lambda} {\bm{H}}_{0}%
^{\prime-1}\right)  \right)  ,$ which is positive definite. Thus, \Cref{J2}
also holds.

\textbf{Part 3:} To show \Cref{Asym1}, we first prove that $LM_{1N}\left(
\hat{\bm{\Lambda}}\right)  -\max_{1\leq j<k\leq S}LM_{0N}\left(
j,k,\bm{\Lambda}\bm{H}_{0}^{\prime-1}\right)  =o_{P}\left(  1\right)  ,$ and
then derive the asymptotic distribution of $\max_{1\leq j<k\leq S}%
LM_{0N}\left(  j,k,\bm{\Lambda}\bm{H}_{0}^{\prime-1}\right)  .$ Similarly, to
show \Cref{Asym2}, we prove that $LM_{2N}\left(  \hat{\bm{\Lambda}}\right)
-\min_{1\leq j<k\leq S}LM_{0N}\left(  j,k,\bm{\Lambda}\bm{H}_{0}^{\prime
-1}\right)  =o_{P}\left(  1\right)  ,$ and derive the asymptotic distribution
of $\min_{1\leq j<k\leq S}LM_{0N}\left(  j,k,\bm{\Lambda} {\bm{H}}_{0}%
^{\prime-1}\right)  .$ We use the fact that $\left\vert LM_{1N}\left(
\hat{\bm{\Lambda }}\right)  -\max_{1\leq j<k\leq S}LM_{0N}\left(
j,k,\bm{\Lambda}\bm{H}_{0}^{\prime-1}\right)  \right\vert $ is bounded by%
\begin{align*}
&  \left\vert LM_{1N}\left(  \hat{\bm{\Lambda }}\right)  -\max_{1\leq j<k\leq
S}LM_{N}\left(  j,k,\bm{\Lambda}\bm{H}_{0}^{\prime-1}\right)  \right\vert \\
&  +\left\vert \max_{1\leq j<k\leq S}LM_{N}\left(
j,k,\bm{\Lambda} {\bm{H}}_{0}^{\prime-1}\right)  -\max_{1\leq j<k\leq
S}LM_{0N}\left(  j,k,\bm{\Lambda}\bm{H}_{0}^{\prime-1}\right)  \right\vert \\
&  \leq\max_{1\leq j<k\leq S}\left\vert LM_{N}\left(  j,k,\hat{\bm{\Lambda }}%
\right)  -LM_{N}\left(  j,k,\bm{\Lambda}\bm{H}_{0}^{\prime-1}\right)
\right\vert \\
&  +\max_{1\leq j<k\leq S}\left\vert LM_{N}\left(  j,k,\bm{\Lambda}\bm{H}_{0}%
^{\prime-1}\right)  -LM_{0N}\left(  j,k,\bm{\Lambda}\bm{H}_{0}^{\prime
-1}\right)  \right\vert \\
&  =o_{P}\left(  1\right)
\end{align*}
and $\left\vert LM_{2N}\left(  \hat{\bm{\Lambda }}\right)  -\min_{1\leq
j<k\leq S}LM_{0N}\left(  j,k,\bm{\Lambda}\bm{H}_{0}^{\prime-1}\right)
\right\vert $ is bounded by%
\begin{align*}
&  \left\vert LM_{2N}\left(  \hat{\bm{\Lambda }}\right)  -\min_{1\leq j<k\leq
S}LM_{N}\left(  j,k,\bm{\Lambda}\bm{H}_{0}^{\prime-1}\right)  \right\vert \\
&  +\left\vert \min_{1\leq j<k\leq S}LM_{N}\left(
j,k,\bm{\Lambda} {\bm{H}}_{0}^{\prime-1}\right)  -\min_{1\leq j<k\leq
S}LM_{0N}\left(  j,k,\bm{\Lambda}\bm{H}_{0}^{\prime-1}\right)  \right\vert \\
&  \leq\max_{1\leq j<k\leq S}\left\vert LM_{N}\left(  j,k,\hat{\bm{\Lambda }}%
\right)  -LM_{N}\left(  j,k,\bm{\Lambda}\bm{H}_{0}^{\prime-1}\right)
\right\vert \\
&  +\max_{1\leq j<k\leq S}\left\vert LM_{N}\left(  j,k,\bm{\Lambda}\bm{H}_{0}%
^{\prime-1}\right)  -LM_{0N}\left(  j,k,\bm{\Lambda}\bm{H}_{0}^{\prime
-1}\right)  \right\vert \\
&  =o_{P}\left(  1\right)  ,
\end{align*}
where the orders in probability come from Part 1 and Part 2. Therefore, by the
asymptotic equivalence lemma, the asymptotic distributions of $LM_{1N}\left(
\hat{\bm{\Lambda }}\right)  $ and $LM_{2N}\left(  \hat{\bm{\Lambda }}\right)
$ are given by those of $\max_{1\leq j<k\leq S}LM_{0N}\left(
j,k,\bm{\Lambda}\bm{H}_{0}^{\prime-1}\right)  $ and $\min_{1\leq j<k\leq
S}LM_{0N}\left(  j,k,\bm{\Lambda}\bm{H}_{0}^{\prime-1}\right)  ,$
respectively$.$

Furthermore, we observe that%
\[
LM_{0N}\left(  j,k,\bm{\Lambda}\bm{H}_{0}^{\prime-1}\right)  =\left(  \pi
_{j}^{-1/2}\bm{V}_{j}-\pi_{k}^{-1/2}\bm{V}_{k}\right)  ^{\prime}\left(
\pi_{j}^{-1}+\pi_{k}^{-1}\right)  ^{-1}\left(  \pi_{j}^{-1/2}\bm{V}_{j}%
-\pi_{k}^{-1/2}\bm{V}_{k}\right)  ,
\]
where $\bm{V}_{j}=\pi_{j}^{1/2}\bm{S}_{0}^{-1/2}\bm{A}_{0}\left(
j,\bm{\Lambda}\bm{H}_{0}^{\prime-1}\right)  .$ Moreover,%
\[
\left(  \mathrm{Var}\left(  \bm{A}_{0}\left(  j,\bm{\Lambda} {\bm{H}}_{0}%
^{\prime-1}\right)  \right)  \right)  ^{-1/2}\bm{A}_{0}\left(
j,\bm{\Lambda}\bm{H}_{0}^{\prime-1}\right)  \overset{d}{\longrightarrow
}N\left(  \bm{0}_{r_{c}\left(  r_{c}+1\right)  /2},\bm{I}_{r_{c}\left(
r_{c}+1\right)  /2}\right)  ,
\]
given \Cref{low_level} \eqref{low_level2}. Therefore,%
\begin{equation}
\bm{V}_{j}\overset{d}{\longrightarrow}\bm{Z}_{j}\sim\mathrm{N}\left(
\bm{0}_{r_{c}\left(  r_{c}+1\right)  /2},\bm{I}_{r_{c}\left(  r_{c}+1\right)
/2}\right)  . \label{asym_v}%
\end{equation}
Since, $\max_{1\leq j<k\leq S}LM_{0N}\left(  j,k,\bm{\Lambda} {\bm{H}}_{0}%
^{\prime-1}\right)  $ and $\min_{1\leq j<k\leq S}LM_{0N}\left(
j,k,\bm{\Lambda}\bm{H}_{0}^{\prime-1}\right)  $ are both continuous functions
of $\bm{V}_{j},$ $j=1,\ldots,S,$ which are independent based on \Cref{Ass1}
(b), we obtain using \Cref{asym_v} that $\max_{1\leq j<k\leq S}LM_{0N}\left(
j,k,\bm{\Lambda}\bm{H}_{0}^{\prime-1}\right)  $ converges in distribution to
\[
\max_{1\leq j<k\leq S}\left(  \left(  \pi_{j}^{-1/2}\bm{Z}_{j}-\pi_{k}%
^{-1/2}\bm{Z}_{k}\right)  ^{\prime}\left(  \pi_{j}^{-1}+\pi_{k}^{-1}\right)
^{-1}\left(  \pi_{j}^{-1/2}\bm{Z}_{j}-\pi_{k}^{-1/2}\bm{Z}_{k}\right)
\right)
\]
and $\min_{1\leq j<k\leq S}LM_{0N}\left(  j,k,\bm{\Lambda}\bm{H}_{0}%
^{\prime-1}\right)  $ converges in distribution to
\[
\min_{1\leq j<k\leq S}\left(  \left(  \pi_{j}^{-1/2}\bm{Z}_{j}-\pi_{k}%
^{-1/2}\bm{Z}_{k}\right)  ^{\prime}\left(  \pi_{j}^{-1}+\pi_{k}^{-1}\right)
^{-1}\left(  \pi_{j}^{-1/2}\bm{Z}_{j}-\pi_{k}^{-1/2}\bm{Z}_{k}\right)
\right)  .
\]
Furthermore, $\pi_{j}^{-1/2}\bm{Z}_{j}-\pi_{k}^{-1/2}\bm{Z}_{k}\sim
\mathrm{N}\left(  \bm{0}_{r_{c}\left(  r_{c}+1\right)  /2},\left(  \pi
_{j}^{-1}+\pi_{k}^{-1}\right)  \bm{I}_{r_{c}\left(  r_{c}+1\right)
/2}\right)  .$
Therefore, we obtain the results in \Cref{Thm1}.
\hfill$\square$\newline

\noindent\textbf{Proof of \Cref{Thm2}}

Consider a pair of groups $\left(  j,k\right)  $. 
We use the decomposition%
\begin{equation}
\frac{1}{N_{j}}\sum_{i=M_{j-1}+1}^{M_{j}}\hat{\bm{\lambda}}_{i}\hat
{\bm{\lambda}}_{i}^{\prime}-\frac{1}{N_{k}}\sum_{i=M_{k-1}+1}^{M_{k}}%
\hat{\bm{\lambda}}_{i}\hat{\bm{\lambda}}_{i}^{\prime}=\bm{R}_{1}(j,k)%
+\bm{R}_{2}(j)-\bm{R}_{2}(k), \label{decomp_alt}%
\end{equation}
where%
\begin{align*}
\bm{R}_{1}(j,k)  &  =\frac{1}{N_{j}}\sum_{i=M_{j-1}+1}^{M_{j}}\hat{\bm{\Xi}}%
^{-1}\bm{\phi}_{i}\bm{\phi}_{i}^{\prime}\hat{\bm{\Xi}}^{\prime-1}-\frac
{1}{N_{k}}\sum_{i=M_{k-1}+1}^{M_{k}}\hat{\bm{\Xi}}^{-1}\bm{\phi}_{i}%
\bm{\phi}_{i}^{\prime}\hat{\bm{\Xi}}^{\prime-1}\\
\bm{R}_{2}(j)  &  =\frac{1}{N_{j}}\sum_{i=M_{j-1}+1}^{M_{j}}\hat{\bm{\lambda}}%
_{i}\hat{\bm{\lambda}}_{i}^{\prime}-\frac{1}{N_{j}}\sum_{i=M_{j-1}+1}^{M_{j}%
}\hat{\bm{\Xi}}^{-1}\bm{\phi}_{i}\bm{\phi}_{i}^{\prime}\hat{\bm{\Xi}}%
^{\prime-1}\\
\bm{R}_{2}(k)  &  =\frac{1}{N_{k}}\sum_{i=M_{k-1}+1}^{M_{k}}\hat{\bm{\lambda}}%
_{i}\hat{\bm{\lambda}}_{i}^{\prime}-\frac{1}{N_{k}}\sum_{i=M_{k-1}+1}^{M_{k}%
}\hat{\bm{\Xi}}^{-1}\bm{\phi}_{i}\bm{\phi}_{i}^{\prime}\hat{\bm{\Xi}}%
^{\prime-1},
\end{align*}
and analyze the limiting behavior of $\bm{R}_{1}(j,k),$ $\bm{R}_{2}(j)$ and
$\bm{R}_{2}(k).$ Starting with $\bm{R}_{2}(j)$, we note that%
\begin{align*}
\bm{R}_{2}(j)  &  =\frac{1}{N_{j}}\sum_{i=M_{j-1}+1}^{M_{j}}\left(
\hat{\bm{\lambda}}_{i}-\hat{\bm{\Xi}}^{-1}\bm{\phi}_{i}\right)  \left(
\hat{\bm{\lambda}}_{i}-\hat{\bm{\Xi}}^{-1}\bm{\phi}_{i}\right)  ^{\prime
}+\frac{1}{N_{j}}\sum_{i=M_{j-1}+1}^{M_{j}}\hat{\bm{\Xi}}^{-1}\bm{\phi}_{i}%
\left(  \hat{\bm{\lambda}}_{i}-\hat{\bm{\Xi}}^{-1}\bm{\phi}_{i}\right)
^{\prime}\\
&  +\frac{1}{N_{j}}\sum_{i=M_{j-1}+1}^{M_{j}}\left(  \hat{\bm{\lambda}}%
_{i}-\hat{\bm{\Xi}}^{-1}\bm{\phi}_{i}\right)  \bm{\phi}_{i}^{\prime}%
\hat{\bm{\Xi}}^{\prime-1}.
\end{align*}
Hence, applying \Cref{A2} \eqref{A21} and \eqref{A22} and noting that
$\hat{\bm{\Xi}}^{-1}=\bm{\Xi}_{0}^{-1}+o_{P}\left(  1\right)  $, we obtain
$\bm{R}_{2}(j)=O_{P}\left(  \delta_{NT}^{-2}\right)  $. Similarly, $\bm{R}_{2}(k)%
=O_{P}\left(  \delta_{NT}^{-2}\right)  .$ In consequence,%
\[
\frac{1}{N_{j}}\sum_{i=M_{j-1}+1}^{M_{j}}\hat{\bm{\lambda}}_{i}\hat
{\bm{\lambda}}_{i}^{\prime}-\frac{1}{N_{k}}\sum_{i=M_{k-1}+1}^{M_{k}}%
\hat{\bm{\lambda}}_{i}\hat{\bm{\lambda}}_{i}^{\prime}=\bm{R}_{1}(j)+O_{P}\left(
\delta_{NT}^{-2}\right)  .
\]
Using the moment conditions in \Cref{Ass3} (b), we have $\bm{R}_{1}(j,k)%
=\bm{R}_{0}(j,k)+o_{P}\left(  1\right)  $, where%
\begin{equation}
\bm{R}_{0}(j,k)=\bm{\Xi}_{0}^{-1}\mathop{\rm plim}_{N\rightarrow\infty}\left(
\frac{1}{N_{j}}\sum_{i=M_{j-1}+1}^{M_{j}}\bm{\phi}_{i}\bm{\phi}_{i}^{\prime
}-\frac{1}{N_{k}}\sum_{i=M_{k-1}+1}^{M_{k}}\bm{\phi}_{i}\bm{\phi}_{i}^{\prime
}\right)  \bm{\Xi}_{0}^{\prime-1}. \label{cond}%
\end{equation}
Hence, for any pair of groups $j$ and $k$,%
\begin{equation}
\frac{1}{N_{j}}\sum_{i=M_{j-1}+1}^{M_{j}}\hat{\bm{\lambda}}_{i}\hat
{\bm{\lambda}}_{i}^{\prime}-\frac{1}{N_{k}}\sum_{i=M_{k-1}+1}^{M_{k}}%
\hat{\bm{\lambda}}_{i}\hat{\bm{\lambda}}_{i}^{\prime}=\bm{R}_{0}(j,k)+o_{P}\left(
1\right)  ,
\label{before_last}
\end{equation}
with $\bm{R}_{0}(j,k)\neq\bm{0}$ when
group-specific heterogeneity emerges in the pair $(j,k)$,
while $\bm{R}_{0}(j,k)=\bm{0}$ when
group-specific heterogeneity does not emerge in the pair $(j,k)$.
Indeed,
for any pair $(j,k)$ with group-specific heterogeneity,
$\bm{R}_{0}(j,k)$ is different from $\bm{0}$ since the rows of $
\bm{\Xi}^{\prime-1}_0$ are linearly independent and $\frac{1}{N_{j}%
}\bm{\Phi}_{j}^{\prime}\bm{\Phi}_{j}-\frac{1}{N_{k}}\bm{\Phi}_{k}^{\prime
}\bm{\Phi}_{k}$ has a nonzero limit in probability.
%
Noting from
\Cref{A2} \eqref{A25} that%
\begin{equation}
\left\Vert \bm{S}\left(  j,k,\hat{\bm{\Lambda}}\right)  -\bm{S}\left(
j,k,\bm{\Phi}\bm{\Xi}_{0}^{\prime-1}\right)  \right\Vert =o_{P}\left(
1\right)  , \label{last}%
\end{equation}
and that $\bm{S}\left(  j,k,\bm{\Phi}\bm{\Xi}_{0}^{-1}\right)  $ has a
positive definite limit in probability according to \Cref{Ass3} (h), we obtain
that $\bm{S}\left(  j,k,\hat{\bm{\Lambda}}\right)  $ has a positive definite limit  in
probability,
equal to
$\bm{S}_0(j,k)$. From \Cref{last,before_last}, there are positive constants $\delta_m, m=1,2,$ such that
\begin{equation}
LM_{mN}\left(  \hat{\bm\Lambda}\right) = N \delta_m +o_P(N), m=1,2,
 \label{last1}%
\end{equation}
with
$$ \delta_1=\max_{1\leq j<k\leq S} \left(\left(\mathrm{Vech}\left(\bm{R}_{0}(j,k)\right)\right)^\prime\left(\bm{S}_0(j,k)\right)^{-1}\mathrm{Vech}\left(\bm{R}_{0}(j,k)\right)\right)$$ and
$$ \delta_2=\min_{1\leq j<k\leq S} \left(\left(\mathrm{Vech}\left(\bm{R}_{0}(j,k)\right)\right)^\prime\left(\bm{S}_0(j,k)\right)^{-1}\mathrm{Vech}\left(\bm{R}_{0}(j,k)\right)\right).$$
Therefore, the results in \Cref{Thm2} follow.\hfill$\square$

\subsection*{Appendix B: Proofs of Results in \Cref{perm_results}
\label{appendix_3}}

\renewcommand{\theequation}{B.\arabic{equation}} \renewcommand{\thesection}{B} \setcounter{equation}{0}

This appendix proves the asymptotic validity of the
permutation approach.

\noindent\textbf{Proof for \Cref{Thm3}}


Suppose $G_{N}$ is a permutation uniformly distributed over $\mathbf G_N$. Let
\[
LM_{0N}\left(  j,k,G_{N}\bm{\Lambda}{\bm{H}}^{\prime-1}_0\right)
=\bm{A}\left(  j,k,G_{N}\bm{\Lambda}{\bm{H}}^{\prime-1}_0\right)  ^{\prime
}\left(  \left(  \frac{1}{\pi_{j}}+\frac{1}{\pi_{k}}\right)  \bm{S}_{0}%
\right)  ^{-1}
\bm{A}\left(  j,k,G_{N}\bm{\Lambda}{\bm{H}}^{\prime-1}_0\right).
\]
Using \Cref{Ass1,Ass2,Ass4}, and proceeding as in the proof of \Cref{Thm1}, we get
$$
\left\vert LM_{1N}\left(  G_{N}\hat{\bm{\Lambda}}\right)  -\max_{1\leq j<k\leq
S}LM_{0N}\left(  j,k,G_{N}\bm{\Lambda}{\bm{H}}^{\prime-1}_0 \right) \right\vert =o_{P}\left(  1\right)
$$
and%
$$
\left\vert LM_{2N}\left(  G_{N}\hat{\bm{\Lambda}}\right)  -\min_{1\leq j<k\leq
S}LM_{0N}\left(  j,k,G_{N}\bm{\Lambda}{\bm{H}}^{\prime-1}_0 \right) \right\vert  =o_{P}\left(  1\right).
$$
Thus, $LM_{1N}\left(  G_{N}\hat{\bm{\Lambda}}\right)$ and $LM_{2N}\left(  G_{N}\hat{\bm{\Lambda}}\right)$
have the same asymptotic limits as\\
$\max_{1\leq j<k\leq S}LM_{0N}\left(  j,k,G_{N}\bm{\Lambda}{\bm{H}}^{\prime-1}_0\right)$ and $\min_{1\leq j<k\leq S} LM_{0N}\left(  j,k,G_{N}\bm{\Lambda}{\bm{H}}^{\prime-1}_0\right)$, respectively.

In order to prove \Cref{Thm3},
we verify the following conditions of \citet*[Theorem 5.1]{CR2013}, written for two permutations $G_{N}$ and $G_{N}^{\prime}$, that are independent, uniformly distributed over $\mathbf G_N$, and independent of $\bm \Lambda$:
\begin{equation}
\left(  T_{1N},U_{1N}\right)  \equiv\left(  \max_{1\leq j<k\leq S}%
LM_{0N}\left(  j,k,G_{N}\bm{\Lambda}{\bm{H}}^{\prime-1}_0\right)  ,\max_{1\leq
j<k\leq S}LM_{0N}\left(  j,G_{N}^{\prime}\bm{\Lambda}{\bm{H}}^{\prime
-1}_0\right)  \right)  \overset{d}{\longrightarrow}\left(  T_{1},U_{1}\right)  ,
\label{CDF_perm1}%
\end{equation}
where $T_{1}=\max_{1\leq l\leq\frac{S\left(  S-1\right)  }{2}}Q_{\frac
{r_{c}\left(  r_{c}+1\right)  }{2}}\left(  l\right)  $ and $U_{1}=\max_{1\leq
l\leq\frac{S\left(  S-1\right)  }{2}}Q_{\frac{r_{c}\left(  r_{c}+1\right)
}{2}}^{\prime}\left(  l\right)  $ are independent with the same CDF
$F_{1}\left(  \cdot\right)  $ and
\begin{equation}
\left(  T_{2N},U_{2N}\right)  \equiv\left(  \min_{1\leq j<k\leq S}%
LM_{0N}\left(  j,k,G_{N}\bm{\Lambda}{\bm{H}}^{\prime-1}_0\right)  ,\min_{1\leq
j<k\leq S}LM_{0N}\left(  j,G_{N}^{\prime}\bm{\Lambda}{\bm{H}}^{\prime
-1}_0\right)  \right)  \overset{d}{\longrightarrow}\left(  T_{2},U_{2}\right)  ,
\label{CDF_perm2}%
\end{equation}
where $T_{2}=\min_{1\leq l\leq\frac{S\left(  S-1\right)  }{2}}Q_{\frac
{r_{c}\left(  r_{c}+1\right)  }{2}}\left(  l\right)  $ and $U_{2}=\min_{1\leq
l\leq\frac{S\left(  S-1\right)  }{2}}Q_{\frac{r_{c}\left(  r_{c}+1\right)
}{2}}^{\prime}\left(  l\right)  $ are independent with the same CDF
$F_{2}\left(  \cdot\right)  $.

We recall that%
\[
\left(  T_{1N},U_{1N}\right)  =\left(  f_{1}\left(  \bm{V}_{N1},\ldots
,\bm{V}_{NS}\right)  ,f_{1}\left(  \bm{W}_{N1},\ldots,\bm{W}_{NS}\right)
\right)  ,
\]
and
\[
\left(  T_{2N},U_{2N}\right)  =\left(  f_{2}\left(  \bm{V}_{N1},\ldots
,\bm{V}_{NS}\right)  ,f_{2}\left(  \bm{W}_{N1},\ldots,\bm{W}_{NS}\right)
\right)  ,
\]
with%
\[
f_{1}\left(  \bm{x}_{1},\ldots,\bm{x}_{S}\right)  =\max_{1\leq j<k\leq
S}\left(  \left(  \pi_{j}^{-1/2}\bm{x}_{j}-\pi_{k}^{-1/2}\bm{x}_{k}\right)
^{\prime}\left(  \pi_{j}^{-1}+\pi_{k}^{-1}\right)  ^{-1}\left(  \pi_{j}%
^{-1/2}\bm{x}_{j}-\pi_{k}^{-1/2}\bm{x}_{k}\right)  \right)  ,
\]
and
\[
f_{2}\left(  \bm{x}_{1},\ldots,\bm{x}_{S}\right)  =\min_{1\leq j<k\leq
S}\left(  \left(  \pi_{j}^{-1/2}\bm{x}_{j}-\pi_{k}^{-1/2}\bm{x}_{k}\right)
^{\prime}\left(  \pi_{j}^{-1}+\pi_{k}^{-1}\right)  ^{-1}\left(  \pi_{j}%
^{-1/2}\bm{x}_{j}-\pi_{k}^{-1/2}\bm{x}_{k}\right)  \right)  ,
\]

\begin{equation}
\bm{V}_{Nj}=\pi_{j}^{1/2}\bm{S}_{0}^{-1/2}\frac{\sqrt{N}}{N_{j}}\sum_{i=1}%
^{N}\iota_{jG_{N}\left(  i\right)  }\mathrm{Vech}\left(  {\bm{H}}%
^{-1}_0\left(  \bm{\lambda}_{c,i}\bm{\lambda}_{c,i}^{\prime}%
-\bm{\Sigma }_{\bm{\Lambda}}\right)  {\bm{H}}^{-1}_0\right)  ,
\end{equation}
and%
\begin{equation}
\bm{W}_{Nj}=\pi_{j}^{1/2}\bm{S}_{0}^{-1/2}\frac{\sqrt{N}}{N_{j}}\sum_{i=1}%
^{N}\iota_{jG_{N}^{\prime}\left(  i\right)  }\mathrm{Vech}\left(
{\bm{H}}^{-1}_0\left(  \bm{\lambda}_{c,i}\bm{\lambda}_{c,i}^{\prime
}-\bm{\Sigma
}_{\bm{\Lambda}}\right)  {\bm{H}}^{-1}_0\right)  .
\end{equation}

In order to show that equations \Cref{CDF_perm1} and \Cref{CDF_perm2} hold,
we show that $\left(  \bm{V}_{N1}^{\prime},\ldots,\bm{V}_{NS}^{\prime
},\bm{W}_{N1}^{\prime},\ldots,\bm{W}_{NS}^{\prime}\right)^\prime  $ is asymptotically
normal.
To this end, we use the Cram\'{e}r-Wold
device and let $\bm{a}$ and $\bm{b}$ be two $\frac{\left(  r_{c}+1\right)
r_{c}}{2}S$-dimensional vectors of real values and prove that%
\begin{equation}
 \Xi^{-1/2}\sum_{j=1}^{S}\left(  \bm{V}_{Nj}^{\prime}\bm{a}_{j}+\bm{W}_{Nj}^{\prime
}\bm{b}_{j}\right)  \overset{d}{\longrightarrow}\mathrm{N}\left(  0, 1 \right)  , \label{normality1}%
\end{equation}
where we define
$ \Xi =\sum_{j=1}^{S}\bm{a}_{j}^{\prime}\bm{a}_{j}+\sum_{j=1}^{S}\bm{b}_{j}^{\prime}%
\bm{b}_{j}
+2\sum_{j=1}^{S}\sum_{k=1}^{S}\pi_{j}^{1/2}\pi_{k}^{1/2}\bm{a}_{j}^{\prime}\bm{b}_{k}.
$
Define $Z_{Nl}=\Xi^{-1/2}\left(  V_{Nl}+W_{Nl}\right)  \ $such that%
\[
V_{Nl}=\sum_{j=1}^{S}\frac{\sqrt{N}}{N_{j}}\pi_{j}^{1/2}\iota_{jG_{N}\left(
l\right)  }\bm{a}_{j}^{\prime}\bm{S}_{0}^{-1/2}\mathrm{Vech}\left(
{\bm{H}}^{-1}_0\left(  \bm{\lambda}_{c,l}\bm{\lambda}_{c,l}^{\prime
}-\bm{\Sigma
}_{\bm{\Lambda}}\right)  {\bm{H}}^{-1}_0\right),
\]%
\[
W_{Nl}=\sum_{j=1}^{S}\frac{\sqrt{N}}{N_{j}}\pi_{j}^{1/2}\iota_{jG_{N}^{\prime
}\left(  l\right)  }\bm{b}_{j}^{\prime}\bm{S}_{0}^{-1/2}\mathrm{Vech}\left(
{\bm{H}}^{-1}_0\left(  \bm{\lambda}_{c,l}\bm{\lambda}_{c,l}^{\prime
}-\bm{\Sigma
}_{\bm{\Lambda}}\right)  {\bm{H}}^{-1}_0\right),
\]
and%
\[
\Xi^{-1/2}\sum_{j=1}^{S}\left(  \bm{V}_{Nj}^{\prime}\bm{a}_{j}+\bm{W}_{Nj}^{\prime
}\bm{b}_{j}\right)= \sum_{l=1}^{N}Z_{Nl}.
\]
Given these notations, we complete the proof of \Cref{Thm3} in three parts.  In the first part, we prove that $\sum_{l=1}^{N}Z_{Nl}$ and $\sum
_{l=1}^{N}\tilde{Z}_{Nl},$ where $\tilde{Z}_{Nl}=Z_{Nl}\mathbb{I}\left(
\sum_{m=1}^{l-1}Z_{Nm}^{2}\leq2\right)  ,$ have the same limiting distribution. In the second part, we establish that $\sum
_{l=1}^{N}\tilde{Z}_{Nl},$ is asymptotically normal with mean $0$ and variance $1$.
In the third part, we deduce \Cref{normality1} and finally show that \Cref{CDF_perm1} and \Cref{CDF_perm2} hold.

\textbf{Part 1:} To prove that $\sum_{l=1}^{N}Z_{Nl}$ and $\sum
_{l=1}^{N}\tilde{Z}_{Nl}$  have the same limiting distribution, we
use \citet*[Theorem 24.2 (iii)]{Davidson1994} by showing that $\sum_{l=1}%
^{N}Z_{Nl}^{2}\overset{P}{\longrightarrow}1$ when $N\rightarrow\infty$. To see
that, we note that%
\[
\sum_{l=1}^{N}Z_{Nl}^{2}=\Xi^{-1}\sum_{l=1}^{N}\left(  V_{Nl}^{2}+W_{Nl}%
^{2}+2V_{Nl}W_{Nl}\right)  .
\]
Hence, it is enough to prove that
\begin{equation}
\sum_{l=1}^{N}V_{Nl}^{2}\overset{P}{\longrightarrow}\sum_{j=1}^{S}%
\bm{a}_{j}^{\prime}\bm{a}_{j},\label{B1}%
\end{equation}
\begin{equation}
\sum_{l=1}^{N}W_{Nl}^{2}\overset{P}{\longrightarrow}\sum_{j=1}^{S}%
\bm{b}_{j}^{\prime}\bm{b}_{j},\label{B2}%
\end{equation}
and
\begin{equation}
\sum_{l=1}^{N}V_{Nl}W_{Nl}\overset{P}{\longrightarrow}\sum_{j=1}^{S}\sum
_{k=1}^{S}\pi_{j}^{1/2}\pi_{k}^{1/2}\bm{a}_{j}^{\prime}\bm{b}_{k}.\label{B3}%
\end{equation}
Note that%
\[
\sum_{l=1}^{N}V_{Nl}^{2}=\sum_{j=1}^{S}\sum_{k=1}^{S}\frac{\pi_{j}^{1/2}%
\pi_{k}^{1/2}N^{2}}{N_{j}N_{k}}\bm{a}_{j}^{\prime}\bm{S}_{0}^{-1/2}\left(
\frac{1}{N}\sum_{l=1}^{N}\iota_{jG_{N}\left(  l\right)  }\iota_{kG_{N}\left(
l\right)  }\bm{B}_{l}\bm{B}_{l}^{\prime}\right)  \bm{S}_{0}^{-1/2}\bm{a}_{k},
\]
where%
\[
\bm{B}_{l}=\mathrm{Vech}\left(  \bm{H}_{0}^{-1}\left(  \bm{\lambda}_{c,l}%
\bm{\lambda}_{c,l}^{\prime}-\bm{\Sigma }_{\bm{\Lambda}}\right)  \bm{H}_{0}%
^{-1}\right)  .
\]
Since $\frac{\pi_{j}^{1/2}\pi_{k}^{1/2}N^{2}}{N_{j}N_{k}}\longrightarrow
\frac{1}{\pi_{j}^{1/2}\pi_{k}^{1/2}},$ we can show that $\sum_{l=1}%
^{N}V_{Nl}^{2}\overset{P}{\longrightarrow}\sum_{j=1}^{S}\bm{a}_{j}^{\prime
}\bm{a}_{j}=\sum_{j=1}^{S}\sum_{k=1}^{S}\bm{a}_{j}^{\prime}\bm{a}_{k}%
\mathbb{I}\left(  j=k\right) $
 if
 %
$
\bm{a}_{j}^{\prime}\bm{S}_{0}^{-1/2}\left(  \frac{1}{N}\sum_{l=1}^{N}%
\iota_{jG_{N}\left(  l\right)  }\iota_{kG_{N}\left(  l\right)  }%
\bm{B}_{l}\bm{B}_{l}^{\prime}\right)  \bm{S}_{0}^{-1/2}\bm{a}_{k}%
\overset{P}{\longrightarrow}\bm{a}_{j}^{\prime}\bm{a}_{k}\pi_{j}^{1/2}\pi
_{k}^{1/2}\mathbb{I}\left(  j=k\right)  .
$
To get the results, we write that $\bm{a}_{j}^{\prime}\bm{S}_{0}%
^{-1/2}\left(  \frac{1}{N}\sum_{l=1}^{N}\iota_{jG_{N}\left(  l\right)  }%
\iota_{kG_{N}\left(  l\right)  }\bm{B}_{l}\bm{B}_{l}^{\prime}\right)
\bm{S}_{0}^{-1/2}\bm{a}_{k}$ is %
$
\bm{a}_{j}^{\prime}\bm{a}_{k}\pi_{j}^{1/2}\pi_{k}^{1/2}\mathbb{I}\left(
j=k\right)  +\frac{1}{N}\sum_{l=1}^{N}\iota_{jG_{N}\left(  l\right)  }%
\iota_{kG_{N}\left(  l\right)  }\bm{a}_{j}^{\prime}\left(  \bm{S}_{0}%
^{-1/2}\bm{B}_{l}\bm{B}_{l}^{\prime}\bm{S}_{0}^{-1/2}-\bm{I}\right)
\bm{a}_{k}+
\bm{a}_{j}^{\prime}\bm{a}_{k}\frac{1}{N}\sum_{l=1}^{N}\left(
\iota_{jG_{N}\left(  l\right)  }\iota_{kG_{N}\left(  l\right)  }-\pi_{j}%
^{1/2}\pi_{k}^{1/2}\mathbb{I}\left(  j=k\right)  \right).
$
So it will be sufficient to
prove that%
\begin{equation}
\frac{1}{N}\sum_{l=1}^{N}\iota_{jG_{N}\left(  l\right)  }\iota_{kG_{N}\left(
l\right)  }\bm{a}_{j}^{\prime}\left(  \bm{S}_{0}^{-1/2}\bm{B}_{l}%
\bm{B}_{l}^{\prime}\bm{S}_{0}^{-1/2}-\bm{I}\right)  \bm{a}_{k}%
\overset{P}{\longrightarrow}0.\label{B111}%
\end{equation}
and
\begin{equation}
\bm{a}_{j}^{\prime}\bm{a}_{k}\frac{1}{N}\sum_{l=1}^{N}\left(  \iota
_{jG_{N}\left(  l\right)  }\iota_{kG_{N}\left(  l\right)  }-\pi_{j}^{1/2}%
\pi_{k}^{1/2}\mathbb{I}\left(  j=k\right)  \right)
\overset{P}{\longrightarrow}0.\label{B112}%
\end{equation}
To see \Cref{B111}, we use $E\left(  \frac{1}{N}\sum_{l=1}^{N}\iota_{jG_{N}\left(  l\right)  }%
\iota_{kG_{N}\left(  l\right)  }\bm{a}_{j}^{\prime}\left(  \bm{S}_{0}%
^{-1/2}\bm{B}_{l}\bm{B}_{l}^{\prime}\bm{S}_{0}^{-1/2}-\bm{I}\right)
\bm{a}_{k}\right)  ^{2}$ equals
\[
\frac{1}{N^{2}}\sum_{l=1}^{N}E\left(  \iota
_{jG_{N}\left(  l\right)  }^{2}\iota_{kG_{N}\left(  l\right)  }^{2}\right)
E\left(  \left(  \bm{a}_{j}^{\prime}\left(  \bm{S}_{0}^{-1/2}\bm{B}_{l}%
\bm{B}_{l}^{\prime}\bm{S}_{0}^{-1/2}-\bm{I}\right)  \bm{a}_{k}\right)
^{2}\right),
\]
where we employ the law of iterated expectation, the independence across $l$ of
$\bm{\lambda}_{c,l}$ conditional on the permutation $G_{N},$ and the
independence of $G_{N}$ and $\bm{\lambda}_{c,l}.$ Because $E\left(
\iota_{jG_{N}\left(  l\right)  }^{2}\iota_{kG_{N}\left(  l\right)  }%
^{2}\right)  =\frac{N_{j}}{N}$ if $j=k$\ and $E\left(  \iota_{jG_{N}\left(
l\right)  }^{2}\iota_{kG_{N}\left(  l\right)  }^{2}\right)  =0$ if $j\neq k,$
and $E\left(  \left(  \bm{a}_{j}^{\prime}\left(  \bm{S}_{0}^{-1/2}%
\bm{B}_{l}\bm{B}_{l}^{\prime}\bm{S}_{0}^{-1/2}-\bm{I}\right)  \bm{a}_{j}%
\right)  ^{2}\right)  \leq C$ given \Cref{Ass1} (b), we have that%
\[
E\left(  \frac{1}{N}\sum_{l=1}^{N}\iota_{jG_{N}\left(  l\right)  }%
\iota_{kG_{N}\left(  l\right)  }\bm{a}_{j}^{\prime}\left(  \bm{S}_{0}%
^{-1/2}\bm{B}_{l}\bm{B}_{l}^{\prime}\bm{S}_{0}^{-1/2}-\bm{I}\right)
\bm{a}_{k}\right)  ^{2}\leq C\frac{N_{j}}{N}\frac{1}{N}
=o\left(  1\right).
\]
Thus, \Cref{B111} holds.\ For \Cref{B112}, we
also observe that\\
$E\left(  \frac{1}{N}\sum_{l=1}^{N}\left(  \iota_{jG_{N}\left(  l\right)
}\iota_{kG_{N}\left(  l\right)  }-\pi_{j}^{1/2}\pi_{k}^{1/2}\mathbb{I}\left(
j=k\right)  \right)  \right)  ^{2}$ equals
\[
\frac{1}{N^{2}}\sum_{l=1}^{N}\sum_{m=1}%
^{N}E\left(  \iota_{jG_{N}\left(  l\right)  }\iota_{jG_{N}\left(  m\right)
}\iota_{kG_{N}\left(  l\right)  }\iota_{kG_{N}\left(  m\right)  }\right)
-2\pi_{j}^{1/2}\pi_{k}^{1/2}\mathbb{I}\left(  j=k\right)  \frac{1}{N} \sum_{l=1}%
^{N}E\left(  \iota_{jG_{N}\left(  l\right)  }\iota_{kG_{N}\left(  l\right)
}\right)  +\mathbb{I}\left(  j=k\right)  \pi_{j}\pi_{k}.
\]
Since
$
\frac{1}{N}\sum_{l=1}^{N}E\left(  \iota_{jG_{N}\left(  l\right)  }%
\iota_{kG_{N}\left(  l\right)  }\right)  =\frac{N_{j}}{N}\mathbb{I}\left(
j=k\right)  \longrightarrow\pi_{j}\mathbb{I}\left(  j=k\right)  ,
$
and
\[
\frac{1}{N^{2}}\sum_{l=1}^{N}\sum_{m=1}^{N}E\left(  \iota_{jG_{N}\left(
l\right)  }\iota_{jG_{N}\left(  m\right)  }\iota_{kG_{N}\left(  l\right)
}\iota_{kG_{N}\left(  m\right)  }\right)
=\frac{\mathbb{I}\left(  j=k\right)}{N^{2}}\sum_{l=1}^{N}%
\sum_{m=1}^{N}E\left(  \iota_{jG_{N}\left(  l\right)  }^{2}\iota
_{jG_{N}\left(  m\right)  }^{2}\right)   \longrightarrow\pi_{j}^{2}\mathbb{I}\left(
j=k\right)  ,
\]
the latter using $E\left(  \iota_{jG_{N}\left(  l\right)  }^{2}\iota
_{jG_{N}\left(  m\right)  }^{2}\right)  =\frac{N_{j}}{N}\frac{N_{j}-1}%
{N-1}$ for $l\ne m$, we have%
\[
E\left(  \frac{1}{N}\sum_{l=1}^{N}\left(  \iota_{jG_{N}\left(  l\right)
}\iota_{kG_{N}\left(  l\right)  }-\pi_{j}^{1/2}\pi_{k}^{1/2}\mathbb{I}\left(
j=k\right)  \right)  \right)  ^{2}=2\pi_{j}^{2}\mathbb{I}\left(  j=k\right)
-2\pi_{j}^{2}\mathbb{I}\left(  j=k\right)   +o\left(  1\right)  =o\left(  1\right)  .
\]
Given \Cref{B111} and \Cref{B112}, \Cref{B1} follows. \Cref{B2} also
follows with identical steps.

Let us now look at \Cref{B3}. Since $\frac{\pi_{j}^{1/2}\pi_{k}^{1/2}N^{2}%
}{N_{j}N_{k}}\longrightarrow\frac{1}{\pi_{j}^{1/2}\pi_{k}^{1/2}}$ and%
\[
\sum_{l=1}^{N}V_{Nl}W_{Nl}=\sum_{j=1}^{S}\sum_{k=1}^{S}\frac{\pi_{j}^{1/2}%
\pi_{k}^{1/2}N^{2}}{N_{j}N_{k}}\bm{a}_{j}^{\prime}\bm{S}_{0}^{-1/2}\left(
\frac{1}{N}\sum_{l=1}^{N}\iota_{jG_{N}\left(  l\right)  }\iota_{kG_{N}%
^{\prime}\left(  l\right)  }\bm{B}_{l}\bm{B}_{l}^{\prime}\right)
\bm{S}_{0}^{-1/2}\bm{b}_{k},
\]
we can obtain that $\sum_{l=1}^{N}V_{Nl}W_{Nl}\overset{P}{\longrightarrow
}\sum_{j=1}^{S}\sum_{k=1}^{S}\pi_{j}^{1/2}\pi_{k}^{1/2}\bm{a}_{j}^{\prime
}\bm{b}_{k}$\ if%
\[
\bm{a}_{j}^{\prime}\bm{S}_{0}^{-1/2}\left(  \frac{1}{N}\sum_{l=1}^{N}%
\iota_{jG_{N}\left(  l\right)  }\iota_{kG_{N}^{\prime}\left(  l\right)
}\bm{B}_{l}\bm{B}_{l}^{\prime}\right)  \bm{S}_{0}^{-1/2}\bm{b}_{k}%
\overset{P}{\longrightarrow}\bm{a}_{j}^{\prime}\bm{b}_{k}\pi_{j}\pi_{k}.
\]
To do so, we decompose $\bm{a}_{j}^{\prime}\bm{S}_{0}^{-1/2}\left(  \frac
{1}{N}\sum_{l=1}^{N}\iota_{jG_{N}\left(  l\right)  }\iota_{kG_{N}^{\prime
}\left(  l\right)  }\bm{B}_{l}\bm{B}_{l}^{\prime}\right)  \bm{S}_{0}%
^{-1/2}\bm{b}_{k}$ as%
\begin{equation}
\bm{a}_{j}^{\prime}\bm{b}_{k}\pi_{j}\pi_{k}+\frac{1}{N}\sum_{l=1}^{N}%
\iota_{jG_{N}\left(  l\right)  }\iota_{kG_{N}^{\prime}\left(  l\right)
}\bm{a}_{j}^{\prime}\left(  \bm{S}_{0}^{-1/2}\bm{B}_{l}\bm{B}_{l}^{\prime
}\bm{S}_{0}^{-1/2}-\bm{I}\right)  \bm{b}_{k}+\bm{a}_{j}^{\prime}%
\bm{b}_{k}\frac{1}{N}\sum_{l=1}^{N}\left(  \iota_{jG_{N}\left(  l\right)
}\iota_{kG_{N}^{\prime}\left(  l\right)  }-\pi_{j}\pi_{k}\right)  ,\label{B31}%
\end{equation}
and show that%
\begin{equation}
\frac{1}{N}\sum_{l=1}^{N}\iota_{jG_{N}\left(  l\right)  }\iota_{kG_{N}%
^{\prime}\left(  l\right)  }\bm{a}_{j}^{\prime}\left(  \bm{S}_{0}%
^{-1/2}\bm{B}_{l}\bm{B}_{l}^{\prime}\bm{S}_{0}^{-1/2}-\bm{I}\right)
\bm{b}_{k}\overset{P}{\longrightarrow}0\label{B311}%
\end{equation}
and
\begin{equation}
\frac{1}{N}\sum_{l=1}^{N}\left(  \iota_{jG_{N}\left(  l\right)  }\iota
_{kG_{N}^{\prime}\left(  l\right)  }-\pi_{j}\pi_{k}\right)
\overset{P}{\longrightarrow}0.\label{B312}%
\end{equation}

To get \Cref{B311}, we use the equality%
\begin{align*}
&E\left(  \frac{1}{N}\sum_{l=1}^{N}\iota_{jG_{N}\left(  l\right)  }%
\iota_{kG_{N}^{\prime}\left(  l\right)  }\bm{a}_{j}^{\prime}\left(
\bm{S}_{0}^{-1/2}\bm{B}_{l}\bm{B}_{l}^{\prime}\bm{S}_{0}^{-1/2}-\bm{I}\right)
\bm{b}_{k}\right)  ^{2}\\
&=\frac{1}{N^{2}}\sum_{l=1}^{N}E\left(  \iota
_{jG_{N}\left(  l\right)  }
\iota_{kG_{N}^{\prime}\left(  l\right)}
\right)  E\left(  \left(  \bm{a}_{j}^{\prime}\left(  \bm{S}_{0}%
^{-1/2}\bm{B}_{l}\bm{B}_{l}^{\prime}\bm{S}_{0}^{-1/2}-\bm{I}\right)
\bm{b}_{k}\right)  ^{2}\right),
\end{align*}
where we apply again the law of iterated expectation, the independence across
$l$ of $\bm{\lambda}_{c,l}$ conditional on the permutations $G_{N}$ and
$G_{N}^{\prime},$ and the independence of $G_{N}$, $G_{N}^{\prime}$, and
$\bm{\lambda}_{c,l}.$ In addition,   $E\left(\left(  \bm{a}_{j}^{\prime}\left(  \bm{S}_{0}^{-1/2}\bm{B}_{l}\bm{B}_{l}%
^{\prime}\bm{S}_{0}^{-1/2}-\bm{I}\right)  \bm{b}_{k}\right)  ^{2}\right)  \leq
C$ given \Cref{Ass1} (b), and
\[
E\left(  \iota_{jG_{N}\left(  l\right)  }
\iota_{kG_{N}^{\prime}\left(  l\right)  }
\right)
=E\left(  \iota_{jG_{N}\left(  l\right)  }
\right)
E\left(  \iota_{kG_{N}^{\prime}\left(
l\right)  }
\right)  =
\frac{N_{j}}{N}
\frac{N_{k}}{N}
\]
due to the independence of $G_{N}$\ and $G_{N}^{\prime}$. As a result, we get%
\begin{align*}
&E\left(  \frac{1}{N}\sum_{l=1}^{N}\iota_{jG_{N}\left(  l\right)  }%
\iota_{kG_{N}^{\prime}\left(  l\right)  }\bm{a}_{j}^{\prime}\left(
\bm{S}_{0}^{-1/2}\bm{B}_{l}\bm{B}_{l}^{\prime}\bm{S}_{0}^{-1/2}-\bm{I}\right)
\bm{b}_{k}\right)  ^{2}\\
&\leq C\frac{1}{N}
\frac{N_{j}}{N}
\frac{N_{k}}{N}
=o\left( 1\right).
\end{align*}

For \Cref{B312}, we noted that $E\left(  \frac{1}{N}\sum_{l=1}^{N}\left(  \iota_{jG_{N}\left(  l\right)
}\iota_{kG_{N}^{\prime}\left(  l\right)  }-\pi_{j}\pi_{k}\right)
\right)  ^{2}$ equals
\[
\frac{1}{N^{2}}\sum_{l=1}^{N}\sum_{m=1}^{N}E\left(
\iota_{jG_{N}\left(  l\right)  }\iota_{jG_{N}\left(  m\right)  }\iota
_{kG_{N}^{\prime}\left(  l\right)  }\iota_{kG_{N}^{\prime}\left(  m\right)
}\right)  -2\pi_{j}\pi_{k}\frac{1}{N}\sum_{l=1}^{N}E\left(  \iota
_{jG_{N}\left(  l\right)  }\iota_{kG_{N}^{\prime}\left(  l\right)  }\right)
+\pi_{j}^{2}\pi_{k}^{2}.
\]
Since %
$
\frac{1}{N}\sum_{l=1}^{N}E\left(  \iota_{jG_{N}\left(  l\right)  }%
\iota_{kG_{N}^{\prime}\left(  l\right)  }\right)  =\frac{1}{N}\sum_{l=1}%
^{N}E\left(  \iota_{jG_{N}\left(  l\right)  }\right)  E\left(  \iota
_{kG_{N}^{\prime}\left(  l\right)  }\right)  =\frac{N_{j}}{N}\frac{N_{k}}%
{N}\longrightarrow\pi_{j}\pi_{k},
$
and%
\begin{align*}
&\frac{1}{N^{2}}\sum_{l=1}^{N}\sum_{m=1}^{N}E\left(  \iota_{jG_{N}\left(
l\right)  }\iota_{jG_{N}\left(  m\right)  }\iota_{kG_{N}^{\prime}\left(
l\right)  }\iota_{kG_{N}^{\prime}\left(  m\right)  }\right)    \\
& =\frac
{1}{N^{2}}\sum_{l=1}^{N}\sum_{m=1}^{N}E\left(  \iota_{jG_{N}\left(  l\right)
}\iota_{jG_{N}\left(  m\right)  }\right)  E\left(  \iota_{kG_{N}^{\prime
}\left(  l\right)  }\iota_{kG_{N}^{\prime}\left(  m\right)  }\right)  \\
& =\frac{1}{N^{2}}\sum_{l=1}^{N}E\left(  \iota_{jG_{N}\left(  l\right)  }%
^{2}\right)  E\left(  \iota_{kG_{N}^{\prime}\left(  l\right)  }^{2}\right)
+\frac{1}{N^{2}}\underset{l\neq m}{\sum_{l=1}^{N}\sum_{m=1}^{N}}E\left(
\iota_{jG_{N}\left(  l\right)  }\iota_{jG_{N}\left(  m\right)  }\right)
E\left(  \iota_{kG_{N}^{\prime}\left(  l\right)  }\iota_{kG_{N}^{\prime
}\left(  m\right)  }\right)  \\
& =\frac{N}{N^{2}}\frac{N_{j}}{N}\frac{N_{k}}{N}+\frac{N\left(  N-1\right)
}{N^{2}}\frac{N_{j}}{N}\frac{N_{j}-1}{N-1}\frac{N_{k}}{N}\frac{N_{k}-1}%
{N-1}\\
& =\frac{N_{j}}{N}\frac{N_{j}-1}{N}\frac{N_{k}}{N}\frac{N_{k}-1}{N-1}+o\left(
1\right) \\
&=\left(  \pi_{j}\pi_{k}\right)  ^{2}+o(1).
\end{align*}
Thus, we have $E\left(  \frac{1}{N}\sum_{l=1}^{N}\left(  \iota_{jG_{N}\left(
l\right)  }\iota_{kG_{N}^{\prime}\left(  l\right)  }-\pi_{j}\pi
_{k}\right)  \right)  ^{2}=o\left(  1\right) $, which implies \Cref{B312}. Hence, we have \Cref{B3}. Since \Cref{B1}, \Cref{B1} and
\Cref{B3} hold, we conlude that $\sum_{l=1}^{N}Z_{Nl}^{2}%
\overset{P}{\longrightarrow}1.$ Therefore, from  \citet*[Theorem 24.2 (iii)]{Davidson1994}, we conclude that $\sum_{l=1}^{N}Z_{Nl}$ and $\sum_{l=1}^{N}\tilde{Z}_{Nl}$
have the same limiting distribution.
\bigskip

\textbf{Part 2:} Second, we verify that $\tilde{Z}_{Nl}$, $1\leq l\leq N$, satisfy the
conditions of  \citet*[Theorem 2.1]{McLeish1974}, implying that $\sum_{l=1}^{N}\tilde{Z}_{Nl}$ is asymptotically $N(0,1)$. In
particular, we have to prove that (a) $\sum_{l=1}^{N}\tilde{Z}_{Nl}%
^{2}\overset{P}{\longrightarrow}1$ as $N\rightarrow\infty,$ (b) $\max_{1\leq
l\leq N}\left\vert \tilde{Z}_{Nl}\right\vert \overset{P}{\longrightarrow}0$ as
$N\longrightarrow\infty,$ (c) for any real $v,$ $E\left(  T_{N}\right)
\longrightarrow1$ as $N\longrightarrow\infty$ and $\left\{  T_{N}\right\}  $
is uniformily integrable, where $T_{N}=%
{\displaystyle\prod\limits_{l=1}^{N}}
\left(  1+iv\tilde{Z}_{Nl}\right)  $, and $i$ denotes the complex number $\sqrt{-1}$ in Part 2 of the proof.


For (a), we use the fact that $\sum_{l=1}^{N}Z_{Nl}^{2}\overset{P}{\longrightarrow}1$ and
 \citet*[Theorem 24.2 (ii)]{Davidson1994}, to conclude that $\sum_{l=1}^{N}\tilde
{Z}_{Nl}^{2}\overset{P}{\longrightarrow}1$.
To prove (c), we define $\tilde{T}_{l-1}=%
{\displaystyle\prod\limits_{m=1}^{l-1}}
\left(  1+iv\tilde{Z}_{Nm}\right)  $
and use the identity%
\begin{equation}
\tilde{T}_{N}=%
{\displaystyle\prod\limits_{l=1}^{N}}
\left(  1+iv\tilde{Z}_{Nl}\right)  =1+iv\sum_{l=1}^{N}\tilde{T}_{l-1}\tilde{Z}%
_{Nl}=1+iv\sum_{l=1}^{N}\tilde{T}_{l-1}Z_{Nl}\mathbb{I}\left(  \sum_{m=1}^{l-1}%
Z_{Nm}^{2}\leq2\right).
\label{TN}
\end{equation}
Furthermore, we have
\begin{align*}
& E\left(  Z_{Nl}|\bm{\lambda}_{c,1},\ldots,\bm{\lambda}_{c,l-1}
\right)  \\
& =\Xi^{-1/2}E\left( {V}_{Nl}|\bm{\lambda}_{c,1}%
,\ldots,\bm{\lambda}_{c,l-1}
\right)  +\Xi^{-1/2}E\left( {W}_{Nl}|\bm{\lambda}_{c,1},\ldots,\bm{\lambda}_{c,l-1}
\right)  \\
& =\Xi^{-1/2}\sum_{j=1}^{S}\bm{a}_{j}^{\prime}\bm{S}_{0}^{-1/2}\pi_{j}%
^{1/2}\frac{\sqrt{N}}{N_{j}} E\iota_{jG_{N}\left(  l\right)  } \left(
\mathrm{Vech}\left(  \bm{H}_{0}^{-1}\left(  \bm{\lambda}_{c,l}%
\bm{\lambda}_{c,l}^{\prime}-\bm{\Sigma }_{\bm{\Lambda}}\right)  \bm{H}_{0}%
^{-1}\right)  |\bm{\lambda}_{c,1},\ldots,\bm{\lambda}_{c,l-1}
\right)
\\
& +\Xi^{-1/2}\sum_{j=1}^{S}\bm{b}_{j}^{\prime}\bm{S}_{0}^{-1/2}\pi_{j}%
^{1/2}\frac{\sqrt{N}}{N_{j}} E\iota_{jG_{N}^{\prime}\left(  l\right)  } \left(
\mathrm{Vech}\left(  \bm{H}_{0}^{-1}\left(  \bm{\lambda}_{c,l}%
\bm{\lambda}_{c,l}^{\prime}-\bm{\Sigma }_{\bm{\Lambda}}\right)  \bm{H}_{0}%
^{-1}\right)  |\bm{\lambda}_{c,1},\ldots,\bm{\lambda}_{c,l-1}
\right)
\\
& =0.
\end{align*}
Combining this result with \Cref{TN}, we get from the law of iterated
expectations that $E\left(  T_{N}\right) = E\left(  \tilde{T}_{N}\right)  =1.$ In addition, $\sup_{N}E\left(
\max_{1\leq l\leq N}Z_{Nl}^{2}\right)  \leq$ $\sup_{N}E\left(  \sum_{l=1}%
^{N}Z_{Nl}^{2}\right)  \leq C<\infty.$ Hence,  \citet*[Theorem 24.2 (i)]{Davidson1994} implies that
$T_{N}$ is uniformly integrable, completing the proof of (c).

For (b), we note that $ \max_{1\leq l\leq N}\left\vert \tilde Z_{Nl}\right\vert \overset{P}{\longrightarrow} 0$ if we can show that
 $ \max_{1\leq l\leq N}\left\vert Z_{Nl}\right\vert \overset{P}{\longrightarrow} 0$.
Consequently, we observe that for any $\varepsilon>0,$ and for some scalar
$\eta>0,$%
\[
0\leq P\left(  \max_{1\leq l\leq N}\left\vert Z_{Nl}\right\vert >\varepsilon
\right)  =P\left(
{\displaystyle\bigcup\limits_{1\leq l\leq N}}
\left\{  \left\vert Z_{Nl}\right\vert >\varepsilon\right\}  \right)  \leq
\sum_{l=1}^{N}P\left(  \left\vert Z_{Nl}\right\vert >\varepsilon\right)
\leq\varepsilon^{-\left(  2+\eta\right)  }\sum_{l=1}^{N}E\left(  \left\vert
Z_{Nl}\right\vert ^{2+\eta}\right)  ,
\]
where the last inequality follows from Markov’s inequality. Therefore,
(b) is satisfied if we prove that as $N\longrightarrow\infty,$%
we have
\[
\sum_{l=1}^{N}E\left(  \left\vert Z_{Nl}\right\vert ^{2+\eta}\right)
\longrightarrow0,
\]
for any $\varepsilon>0$ and for some scalar $\eta>0.$ Using the c-r
inequality, we have%
\begin{align*}
E\left(  \left\vert Z_{Nl}\right\vert ^{2+\eta}\right)     =E\left(
\left\vert \Xi^{-1/2} V_{Nl}+\Xi^{-1/2} W_{Nl}\right\vert ^{2+\eta}\right)   \leq2^{1+\eta}E\left(  \left\vert \Xi^{-1/2}{V}_{Nl}\right\vert ^{2+\eta} \right)
+2^{1+\eta}E\left(  \left\vert
\Xi^{-1/2}{W}_{Nl}\right\vert  ^{2+\eta} \right).
\end{align*}
We note that $E\left(  \left\vert \Xi^{-1/2}{V}_{Nl}
\right\vert  ^{2+\eta}\ \right)$
can be expressed as
\[
\Xi^{-\left(  2+\eta\right)  /2}S^{2+\eta}E\left(
\left\vert \frac{1}{S}\sum_{j=1}^{S}\frac{N^{1/2}\pi_{j}^{1/2}}{N_{j}}\bm{a}_{j}^{\prime
}\bm{S}_{0}^{-1/2}\iota_{jG_{N}\left(  l\right)  }\left(  \mathrm{Vech}\left(
\bm{H}_{0}^{-1}\left(  \bm{\lambda}_{c,l}\bm{\lambda}_{c,l}^{\prime
}-\bm{\Sigma }_{\bm{\Lambda}}\right)  \bm{H}_{0}^{-1}\right)  \right)
\right\vert  ^{2+\eta} \right).
\]
Using the Jensen inequality, the Cauchy-Schwarz inequality, and the
independence of $G_{N}$ and the factor loadings, we find that $E\left(  \left\vert \Xi^{-1/2}%
{V}_{Nl}\right\vert ^{2+\eta} \right)$ is bounded from above by%
\begin{align*}
& \Xi^{-\left(  2+\eta\right)  /2}S^{1+\eta}\sum_{j=1}^{S}E\left(  \left\vert
\frac{N^{1/2}\pi_{j}^{1/2}}{N_{j}}\bm{a}_{j}^{\prime}\bm{S}_{0}^{-1/2}\iota_{jG_{N}\left(
l\right)  }\left(  \mathrm{Vech}\left(  \bm{H}_{0}^{-1}\left(
\bm{\lambda}_{c,l}\bm{\lambda}_{c,l}^{\prime}-\bm{\Sigma }_{\bm{\Lambda}}%
\right)  \bm{H}_{0}^{-1}\right)  \right)  \right\vert ^{2+\eta}\right)  \\
& \leq\Xi^{-\left(  2+\eta\right)  /2}S^{1+\eta}\sum_{j=1}^{S}E\left(
\left\Vert \frac{N^{1/2}\pi_{j}^{1/2}}{N_{j}}\bm{a}_{j}^{\prime}\bm{S}_{0}^{-1/2}%
\iota_{jG_{N}\left(  l\right)  }\right\Vert ^{2+\eta}\right)  E\left(
\left\Vert \mathrm{Vech}\left(  \bm{H}_{0}^{-1}\left(  \bm{\lambda}_{c,l}%
\bm{\lambda}_{c,l}^{\prime}-\bm{\Sigma }_{\bm{\Lambda}}\right)  \bm{H}_{0}%
^{-1}\right)  \right\Vert ^{2+\eta}\right)  .
\end{align*}
In addition,%
\[
E\left(  \left\Vert \frac{N^{1/2}\pi_{j}^{1/2}}{N_{j}}\bm{a}_{j}^{\prime}\bm{S}_{0}%
^{-1/2}\iota_{jG_{N}\left(  l\right)  }\right\Vert ^{2+\eta}\right)
=\left\Vert \pi_{j}^{1/2} \bm{a}_{j}^{\prime}\bm{S}_{0}^{-1/2}\right\Vert ^{2+\eta}\left(
\frac{N^{1/2}}{N_{j}}\right)  ^{2+\eta}E\left(  \iota_{jG_{N}\left(  l\right)
}^{2+\eta}\right)  \leq C\left(  \frac{N^{\eta/2}}{N_{j}^{1+\eta}}\right)
\]
and %
$
E\left(  \left\Vert \mathrm{Vech}\left(  \bm{H}_{0}^{-1}\left(
\bm{\lambda}_{c,l}\bm{\lambda}_{c,l}^{\prime}-\bm{\Sigma }_{\bm{\Lambda}}%
\right)  \bm{H}_{0}^{-1}\right)  \right\Vert ^{2+\eta}\right)  \leq C,
$
using \Cref{Ass1} (b).
Therefore, $E\left(  \left\vert \Xi^{-1/2}{V}_{Nl}%
\right\vert ^{2+\eta}\right) \leq\Xi^{-\left(  2+\eta\right)
/2}S^{1+\eta}C^{2}\sum_{j=1}^{S}\left(  \frac{N^{\eta/2}}{N_{j}^{1+\eta}}\right)  .$
A similar inequality can be obtained for $E\left(  \left\vert \Xi^{-1/2}{W}_{Nl}\right\vert  ^{2+\eta} \right)$.
We finally have
$$
\sum_{l=1}^{N}E\left(  \left\vert Z_{Nl}\right\vert ^{2+\eta}\right)
\leq2^{2+\eta}\Xi^{-\left(  2+\eta\right)  /2}S^{1+\eta}C^{2}\sum_{j=1}^{S}\left(
\frac{N^{1+\eta/2}}{N_{j}^{1+\eta}}\right)  =O\left(  \frac{1}{N^{\eta/2}%
}\right)  =o\left(  1\right)  .
$$
\bigskip

\textbf{Part 3:} We showed in Part 1 that $\sum_{l=1}^{N}Z_{Nl}$ and
$\sum_{l=1}^{N}\tilde{Z}_{Nl}$ have the same limiting distribution, and proved in Part 2 that $\sum_{l=1}^{N}\tilde{Z}_{Nl}$\ is asymptotically
$\mathrm{N}\left(  0,1\right) $. It follows that%
\[
\sum_{j=1}^{S}\left(  \bm{V}_{Nj}^{\prime}\bm{a}_{j}+\bm{W}_{Nj}^{\prime
}\bm{b}_{j}\right)  \overset{d}{\longrightarrow}\mathrm{N}\left(
0,\Xi\right)  .
\]
Consequently, we deduce that the limit in distribution of $\left(  \bm{V}_{N1}^{\prime},\ldots,\bm{V}_{NS}^{\prime
},\bm{W}_{N1}^{\prime},\ldots,\bm{W}_{NS}^{\prime}\right)^{\prime}  $, denoted
$\left(  \bm{V}_{1}^{\prime},\ldots,\bm{V}_{S}^{\prime},\bm{W}_{1}^{\prime
},\ldots,\bm{W}_{S}^{\prime}\right)^{\prime}$, is multivariate normal with mean $\bm 0$ and covariance matrix such that
\begin{align}
Cov\left(  \bm{V}_{j},\bm{V}_{k}\right)   &  =\bm{0}\text{ if }j\neq k,\\
Cov\left(  \bm{W}_{j},\bm{W}_{k}\right)   &  =\bm{0}\text{ if }j\neq k,\\
Cov\left(  \bm{V}_{j},\bm{W}_{k}\right)   &  =\pi_{j}^{1/2}\pi_{k}%
^{1/2}\bm{I},\\
Var\left(  \bm{V}_{j}\right)   &  =\bm{I},\\
Var\left(  \bm{W}_{j}\right)   &  =\bm{I}.
\end{align}
As a result, we have $\left(  T_{1N},U_{1N}\right) \overset{d}{\longrightarrow}\left(  T_{1},U_{1}\right)$,
and $\left(  T_{2N},U_{2N}\right) \overset{d}{\longrightarrow}\left(  T_{2},U_{2}\right)$.

We still need to show the independence of $T_{1}$ and $U_{1}$ and of $T_{2}$ and $U_{2}$. For this purpose, we write
\[
\left(  T_{1},U_{1}\right)  =\left(  g_{1}\left( \bm{y}_{1,2},\ldots,\bm{y}_{S-1,S} \right) ,
g_{1}\left(\bm{z}_{1,2},\ldots,\bm{z}_{S-1,S}\right) \right)   ,
\]
where%
\[
\bm{y}_{j,k}=\pi
_{j}^{-1/2}\bm{V}_{j}-\pi_{k}^{-1/2}\bm{V}_{k},\text{ \ }j=1,\ldots
,S-1,k=j+1,\ldots,S,
\]%
\[
\bm{z}_{j,k}=\pi_{j}^{-1/2}\bm{W}_{j}-\pi_{k}^{-1/2}\bm{W}_{k},\text{ \ }j=1,\ldots
,S-1,k=j+1,\ldots,S,
\]
and%
\[
g_{1}\left(  \bm{x}_{1,2},\ldots,\bm{x}_{S-1,S}\right)
=\max_{1\leq j<k\leq S}\left(  \left(  \bm{x}_{j,k}\right)
^{\prime}\left(  \pi_{j}^{-1}+\pi_{k}^{-1}\right)  ^{-1}\bm{x}_{j,k}\right)  .
\]
From the normality of $\left(  \bm{V}_{1}^{\prime},\ldots,\bm{V}_{S}^{\prime},\bm{W}_{1}^{\prime},\ldots,\bm{W}_{S}^{\prime}\right)^{\prime}$, it follows through a linear transformation that $\left(  \bm{y}_{1,2}^{\prime},\ldots,\bm{y}_{S-1,S}^{\prime},
\bm{z}_{1,2}^{\prime},\ldots,\bm{z}_{S-1,S}^{\prime}\right)^{\prime}$
is jointly normal. Therefore, to show that $T_{1}$ and $U_{1}$  are independent, it is sufficient to show that
$Cov\left(  \bm{y}_{j_{1},k_{1}},\bm{z}_{j_{2},k_{2} }\right)=0$, for any $j_{1}, k_{1},j_{2}$ and $k_{2} $. We can express $Cov\left(  \bm{y}_{j_{1},k_{1}},\bm{z}_{j_{2},k_{2} }\right)$ as
$$\pi_{j_1}^{-\frac{1}{2}}\pi_{j_2}^{-\frac{1}{2}}Cov\left(  \bm{V}_{j_{1}},\bm{W}_{j_{2}}\right)-\pi_{j_1}^{-\frac{1}{2}}\pi_{k_2}^{-\frac{1}{2}}Cov\left(  \bm{V}_{j_{1}},\bm{W}_{k_{2}}\right)-\pi_{k_1}^{-\frac{1}{2}}\pi_{j_2}^{-\frac{1}{2}}Cov\left(  \bm{V}_{k_1},\bm{W}_{j_{2}}\right)+\pi_{k_1}^{-\frac{1}{2}}\pi_{k_2}^{-\frac{1}{2}}Cov\left(  \bm{V}_{k_{1}},\bm{W}_{ k_{2}}\right),$$
which equals $0$ due to the fact that
$Cov\left(  \bm{V}_{j},\bm{W}_{ k}\right)=\pi_{j}^{\frac{1}{2}}\pi_{k}^{\frac{1}{2}}$, for any $j$ and $k$.
The independence of $T_{2}$ and $U_{2}$ follows using similar steps.
Therefore, equations \Cref{CDF_perm1} and \Cref{CDF_perm2} hold.
\bibliographystyle{chicago}
\bibliography{biblio_DS}

\begin{thebibliography}{}

\bibitem[\protect\citeauthoryear{Ando and Bai}{Ando and Bai}{2017}]{AB2017}
Ando, T. and J.~Bai (2017).
\newblock {Clustering Huge Number of Financial Time Series: A Panel Data
  Approach With High-Dimensional Predictors and Factor Structures}.
\newblock {\em Journal of the American Statistical Association\/}~{\em
  112\/}(519), 1182 -- 1198.

\bibitem[\protect\citeauthoryear{Bai}{Bai}{2003}]{Bai2003}
Bai, J. (2003).
\newblock Inferential theory for factor models of large dimensions.
\newblock {\em Econometrica\/}~{\em 71\/}(1), 135--171.

\bibitem[\protect\citeauthoryear{Bai and Ng}{Bai and Ng}{2002}]{BN2002}
Bai, J. and S.~Ng (2002).
\newblock Determining the number of factors in approximate factor models.
\newblock {\em Econometrica\/}~{\em 70\/}(1), 191--221.

\bibitem[\protect\citeauthoryear{Bai and Ng}{Bai and Ng}{2006}]{BN2006}
Bai, J. and S.~Ng (2006).
\newblock Confidence intervals for diffusion index forecasts and inference for
  factor-augmented regressions.
\newblock {\em Econometrica\/}~{\em 74\/}(4), 1133--1150.

\bibitem[\protect\citeauthoryear{Canay, Romano, and Shaikh}{Canay
  et~al.}{2017}]{CRS2017}
Canay, I.~A., J.~P. Romano, and A.~M. Shaikh (2017).
\newblock {Randomization Tests Under an Approximate Symmetry Assumption}.
\newblock {\em Econometrica\/}~{\em 85}, 1013--1030.

\bibitem[\protect\citeauthoryear{Chung and Romano}{Chung and
  Romano}{2013}]{CR2013}
Chung, E. and J.~P. Romano (2013).
\newblock Exact and asymptotically robust permutation tests.
\newblock {\em Ann. Statist.\/}~{\em 41\/}(2), 484--507.

\bibitem[\protect\citeauthoryear{Davidson}{Davidson}{1994}]{Davidson1994}
Davidson, J. (1994).
\newblock {\em Stochastic Limit Theory: An Introduction for Econometricians}.
\newblock Oxford University Press.

\bibitem[\protect\citeauthoryear{Djogbenou, Gon{\c c}alves, and
  Perron}{Djogbenou et~al.}{2015}]{DGP2015}
Djogbenou, A., S.~Gon{\c c}alves, and B.~Perron (2015).
\newblock Bootstrap inference in regressions with estimated factors and serial
  correlation.
\newblock {\em Journal of Time Series Analysis\/}~{\em 36\/}(3), 481--502.

\bibitem[\protect\citeauthoryear{Djogbenou}{Djogbenou}{2020}]{Djogbenou2019}
Djogbenou, A.~A. (2020).
\newblock {Comovements in the Real Activity of Developed and Emerging
  Economies: A Test of Global versus Specific International Factors}.
\newblock {\em Journal of Applied Econometrics\/}~{\em 35\/}(3), 344--370.

\bibitem[\protect\citeauthoryear{Djogbenou}{Djogbenou}{2021}]{Djogbenou2019b}
Djogbenou, A.~A. (2021).
\newblock Model selection in factor-augmented regressions with estimated
  factors.
\newblock {\em Econometric Reviews\/}~{\em 40\/}(5), 470--503.

\bibitem[\protect\citeauthoryear{Fisher}{Fisher}{1936}]{Fisher1936}
Fisher, R.~A. (1936).
\newblock ``the coefficient of racial likeness'' and the future of craniometry.
\newblock {\em The Journal of the Royal Anthropological Institute of Great
  Britain and Ireland\/}~{\em 66}, 57--63.

\bibitem[\protect\citeauthoryear{Hemerik and Goeman}{Hemerik and
  Goeman}{2018}]{HG2018}
Hemerik, J. and J.~Goeman (2018).
\newblock {Exact testing with random permutations}.
\newblock {\em TEST: An Official Journal of the Spanish Society of Statistics
  and Operations Research\/}~{\em 27\/}(4), 811--825.

\bibitem[\protect\citeauthoryear{Hrazdil and Scott}{Hrazdil and
  Scott}{2013}]{HS2013}
Hrazdil, K. and T.~Scott (2013).
\newblock {The role of industry classification in estimating discretionary
  accruals}.
\newblock {\em Review of Quantitative Finance and Accounting\/}~{\em 40\/}(1),
  15--39.

\bibitem[\protect\citeauthoryear{Kahle and Walkling}{Kahle and
  Walkling}{1996}]{KW1996}
Kahle, K.~M. and R.~A. Walkling (1996).
\newblock The impact of industry classifications on financial research.
\newblock {\em The Journal of Financial and Quantitative Analysis\/}~{\em
  31\/}(3), 309--335.

\bibitem[\protect\citeauthoryear{Kose, Otrok, and Prasad}{Kose
  et~al.}{2012}]{KOP2012}
Kose, M.~A., C.~M. Otrok, and E.~S. Prasad (2012).
\newblock {Global business cycles: convergence or decoupling?}
\newblock {\em International Economic Review\/}~{\em 53\/}(2), 511--538.

\bibitem[\protect\citeauthoryear{Lehmann and Romano}{Lehmann and
  Romano}{2005}]{LR2005}
Lehmann, E.~L. and J.~P. Romano (2005).
\newblock Testing statistical hypotheses.
\newblock {\em Springer Science \& Business Medias\/}.

\bibitem[\protect\citeauthoryear{McLeish}{McLeish}{1974}]{McLeish1974}
McLeish, D.~L. (1974).
\newblock {Dependent Central Limit Theorems and Invariance Principles}.
\newblock {\em The Annals of Probability\/}~{\em 2\/}(4), 620 -- 628.

\bibitem[\protect\citeauthoryear{Negro and Brooks}{Negro and
  Brooks}{2005}]{NB2005}
Negro, M.~D. and R.~Brooks (2005).
\newblock {A Latent Factor Model with Global, Country, and Industry Shocks for
  International Stock Returns}.
\newblock IMF Working Papers 2005/052, International Monetary Fund.

\bibitem[\protect\citeauthoryear{Romano}{Romano}{1990}]{Romano1990}
Romano, J.~P. (1990).
\newblock On the behavior of randomization tests without a group invariance
  assumption.
\newblock {\em Journal of the American Statistical Association\/}~{\em
  85\/}(411), 686--692.

\bibitem[\protect\citeauthoryear{Stock and Watson}{Stock and
  Watson}{2002}]{SW2002}
Stock, J. and M.~Watson (2002).
\newblock Forecasting using principal components from a large number of
  predictors.
\newblock {\em Journal of the American Statistical Association\/}~{\em 97},
  1167--1179.

\bibitem[\protect\citeauthoryear{Sz\'ekely and Rizzo}{Sz\'ekely and
  Rizzo}{2004}]{SR2004}
Sz\'ekely, G.~J. and M.~L. Rizzo (2004).
\newblock Testing for equal distributions in high dimensions.
\newblock {\em InterStat\/}.

\end{thebibliography}

\end{document}